\documentclass[letterpaper,twocolumn,10pt]{article}
\usepackage{usenix}

\usepackage{todonotes}

\usepackage{xspace}
\usepackage{booktabs}
\usepackage{algorithm}
\usepackage{algorithmic}
\usepackage{amsmath,amsfonts}
\usepackage{mathtools}

\usepackage{makecell}
\usepackage{graphicx}
\usepackage{subcaption}
\usepackage{caption}
\usepackage{adjustbox}
\usepackage{balance}
\usepackage{multirow}
\usepackage{hyperref}

\usepackage{tikz}

\newcommand{\name}{\mbox{\texttt{LiDARFace}}\xspace}

\begin{document}

\date{}

\title{\Large \bf Face in the Beam: Exposing and Quantifying Identity Leakage \\ from Automotive LiDAR Scans}

\author{
{\rm Fengxu Yang}\\
ShanghaiTech University\\
yangfx@shanghaitech.edu.cn
\and
{\rm Zaizhou Yang}\\
ShanghaiTech University\\
yangzzh2025@shanghaitech.edu.cn
\and
{\rm Yihui Yan}\\
Shanghai Maritime University\\
yhyan@shmtu.edu.cn
\and
{\rm Zhice Yang}\\
ShanghaiTech University\\
yangzhc@shanghaitech.edu.cn
} 

\maketitle

\begin{abstract}
Vehicle-mounted LiDAR is widely deployed for 3D perception in autonomous driving and ADAS. Although LiDAR does not capture RGB texture, increasing sensing resolution raises an underexplored privacy concern: can sparse facial geometry captured by automotive LiDAR be linked to identity? We study this risk as gallery-based depth-to-RGB linkage, where an adversary associates LiDAR-derived facial observations with identities in an external RGB face gallery. We present \name, a low-barrier measurement pipeline that synthesizes sensor-grounded facial LiDAR observations, reconstructs denser facial depth representations, and performs teacher-guided depth-to-RGB identification by aligning LiDAR-derived embeddings with a frozen RGB face embedding space. We further use ID-aware re-ranking to quantify leakage amplification from multiple observations or gallery images. Experiments on real-person LiDAR captures and controlled synthetic data show that sparse facial geometry enables above-chance identity linkage. \name achieves Rank-1 accuracy up to 30.5\% and Lift@1 up to 156$\times$ over matched random guessing. These results show that raw or lightly processed automotive LiDAR point clouds can expose linkable biometric information, motivating face-region filtering, controlled retention, and access auditing.

\end{abstract}

\section{Introduction}
\label{sec:introduction}

Light detection and ranging~(LiDAR) has become a core perception sensor in autonomous driving and advanced driver-assistance systems~(ADAS), supporting 3D scene understanding for intelligent vehicles~\cite{joseph2021autonomous} and smart transportation infrastructure~\cite{tang2022environmental}. By emitting laser pulses and measuring time-of-flight~(ToF), automotive LiDAR captures 3D point clouds that remain effective under low light, strong illumination, and adverse weather conditions where cameras may degrade~\cite{cheng2014robust,goberville2020analysis}. Consequently, LiDAR-based object detection, tracking, and behavior analysis have become central components of modern intelligent transportation systems~\cite{aung2024review}.

\begin{figure}[t]
    \centering
    \includegraphics[width=0.90\linewidth]{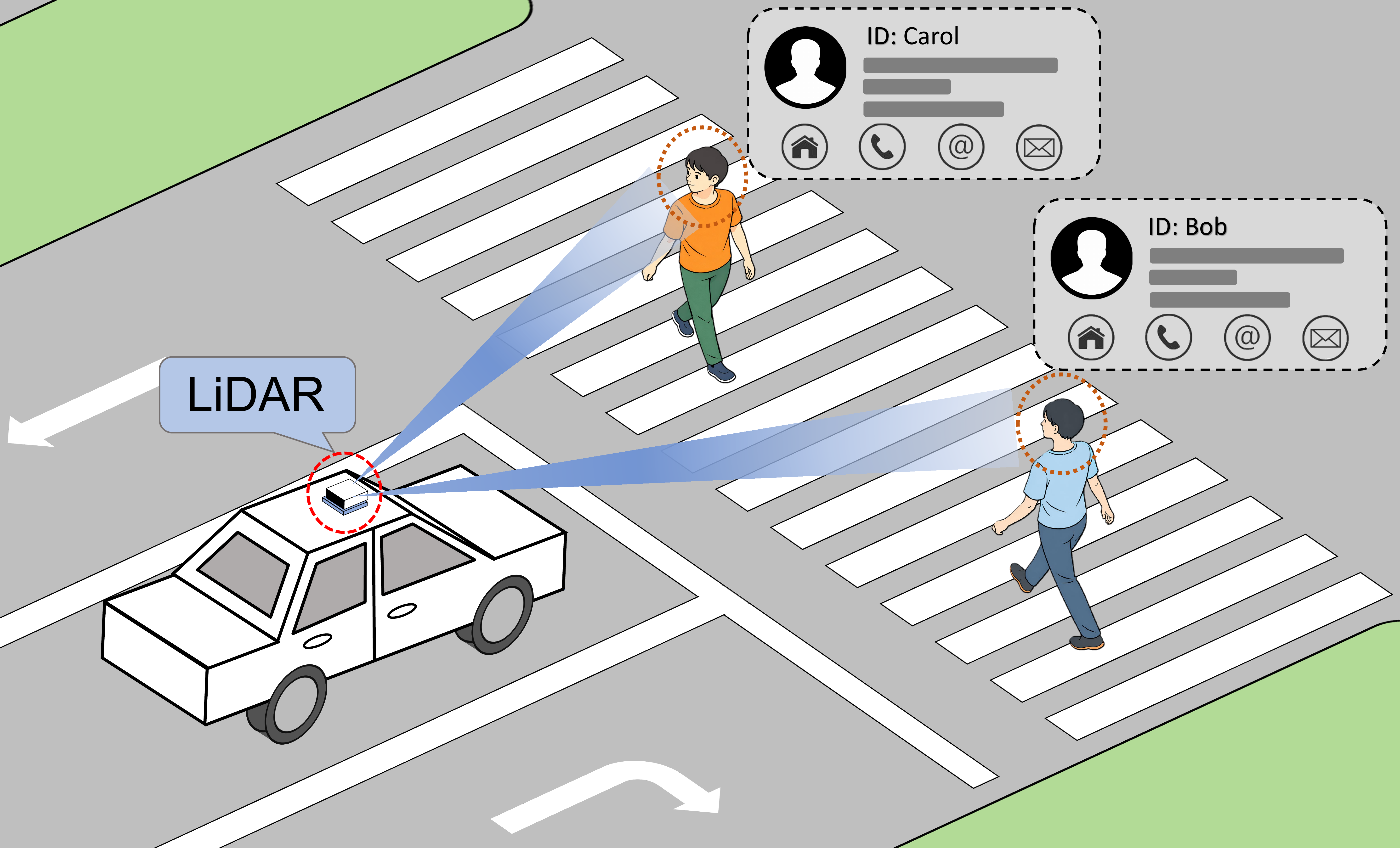}
    \caption{Identity leakage from vehicle-mounted LiDAR in public traffic scenes. Sparse facial geometry captured during routine LiDAR perception may be linked to identities in an external RGB face gallery, even when no co-captured RGB image of the same scene is available.}
    \label{fig:teaser}
\end{figure}

As automotive LiDAR sensors improve in range and angular resolution, they increasingly capture fine-grained geometric information about nearby road users. Commercial vehicle-mounted LiDARs, such as RoboSense RS-LiDAR-M1~\cite{M1datasheet}, still produce point clouds that are much sparser than megapixel RGB images. Nevertheless, when a pedestrian is observed at a favorable distance and orientation, the point cloud may contain structured returns from the head and parts of the facial region. This raises an underexplored privacy question: even without color or texture, can sparse 3D facial geometry expose identity-relevant information? Fig.~\ref{fig:teaser} illustrates this risk scenario.

Vehicle-mounted cameras are widely recognized as privacy-sensitive because they directly record appearance cues such as faces, clothing, and license plates. LiDAR is often perceived as less invasive because it does not record RGB texture. However, a face is also a geometric biometric: facial contour, nasal structure, and other stable mid-face geometry may remain partially observable even in sparse LiDAR returns. If such geometry can be mapped into a representation that supports identity matching, LiDAR data collected for safety or perception may also enable identity linkage.

A practical misuse pathway does not require a dedicated LiDAR face database. Public RGB face images are abundant and can be collected from social media, web photos, or existing face datasets. In contrast, large-scale LiDAR face galleries are scarce and costly to build. An adversary is therefore more likely to use RGB images as external identity references and attempt cross-modal linkage from LiDAR-derived facial geometry to RGB identities. Such linkage matters because identifying a pedestrian is not only a single recognition event: repeated observations may support profiling across time and locations, and an identity link can serve as a pivot to additional public information online. Motivated by this concern, we ask:

\begin{quote}
\emph{Can sparse facial point clouds captured by vehicle-mounted LiDAR be linked to identities in an external RGB gallery?}
\end{quote}

Answering this question is challenging for three reasons. First, public driving datasets rarely contain face-resolved automotive LiDAR scans with identity labels and corresponding RGB references. Second, automotive LiDAR facial observations are sparse, noisy, and incomplete, making the identity signal substantially weaker than in dedicated 3D face-scanning settings. Third, identity linkage must bridge a modality gap between LiDAR-derived geometry and RGB appearance, without assuming paired LiDAR--RGB training data for the target identities.

We address these challenges with \name, a measurement pipeline for evaluating LiDAR identity leakage under the stated threat model. \name is not designed to advance face recognition performance; instead, it instantiates a low-barrier depth-to-RGB linkage pipeline built from accessible data and standard components. At a high level, \name synthesizes sensor-grounded facial LiDAR observations from RGB identity references, applies facial depth completion as a capability amplifier, and trains a depth encoder to align LiDAR-derived depth representations with a frozen RGB face embedding space. It also includes an optional ID-aware re-ranking module to measure how identity-level aggregation can amplify leakage when multiple observations or multiple gallery images are available.

Our evaluation is real-capture-centered. We first assess whether real facial LiDAR captures from a vehicle-mounted RoboSense M1P can be linked to RGB identity references under seen-ID and unseen-ID protocols. We then evaluate whether ID-aware aggregation amplifies leakage from real captures without model retraining. Finally, because real-person capture is difficult to scale and does not readily support controlled factor isolation, we use sensor-grounded synthetic data for large-scale measurements and sensitivity analysis. This structure treats real-person captures as the primary empirical check, while using synthetic data to study scale and factors such as point density, completion method, gallery composition, and the number of observations.

Prior LiDAR-based person re-identification studies primarily focus on body-level cues, such as body shape and gait, often in indoor or short-range settings~\cite{guo2024lidar}. Their goal is typically to improve recognition accuracy from full-body point clouds. In contrast, we study a different privacy channel: face-centric identity linkage from sparse automotive LiDAR facial geometry to external RGB identity references. This setting is cross-modal, gallery-based, and focused on privacy measurement rather than recognition system design. Our main contributions are:

\begin{itemize}
    \item We formulate LiDAR identity leakage in vehicle sensors as gallery-based depth-to-RGB identity linkage from sparse facial geometry, and define the corresponding privacy threat and attacker assumptions.

    \item We design and implement \name, a low-barrier measurement pipeline that combines sensor-grounded facial LiDAR synthesis, facial depth completion, teacher-guided depth-to-RGB identification, and optional ID-aware re-ranking.

    \item We evaluate \name using both real-person LiDAR captures and controlled synthetic data. The real-capture results provide empirical evidence of above-chance identity linkage, while the synthetic experiments enable large-scale measurement and controlled analysis of factors that affect leakage severity.
\end{itemize}
\section{Background}\label{sec:background}

\subsection{Vehicle-mounted LiDAR}

\begin{figure}[t]
    \begin{subfigure}[t]{0.5\linewidth}
        \centering
        \includegraphics[width=\linewidth]{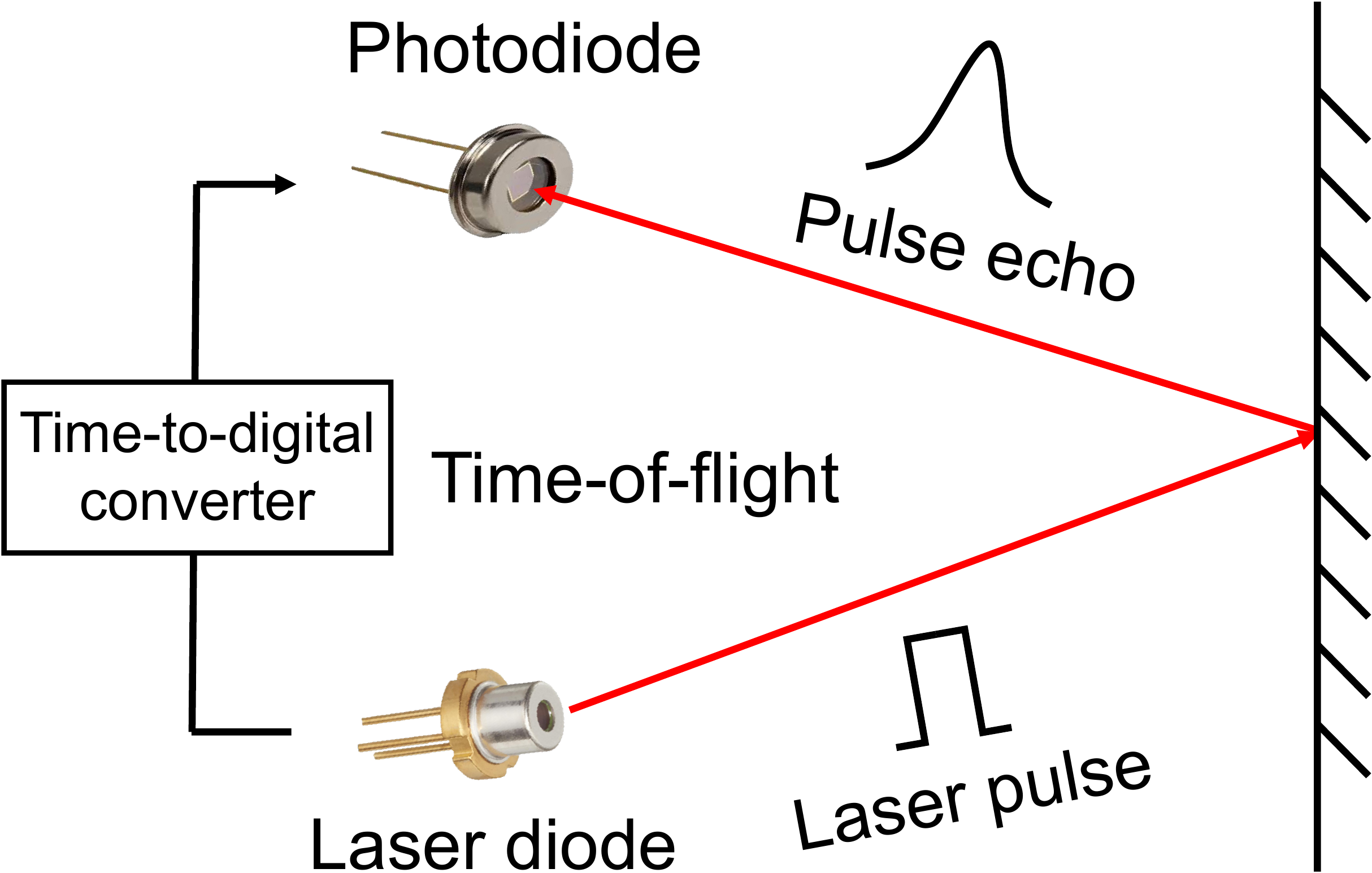}
        \subcaption{Ranging principle.}
        \label{fig:lidar_ranging_principle}
    \end{subfigure}
    \hspace{0.02\linewidth}
    \begin{subfigure}[t]{0.45\linewidth}
        \centering
        \raisebox{3mm}{
        \includegraphics[width=0.9\linewidth]{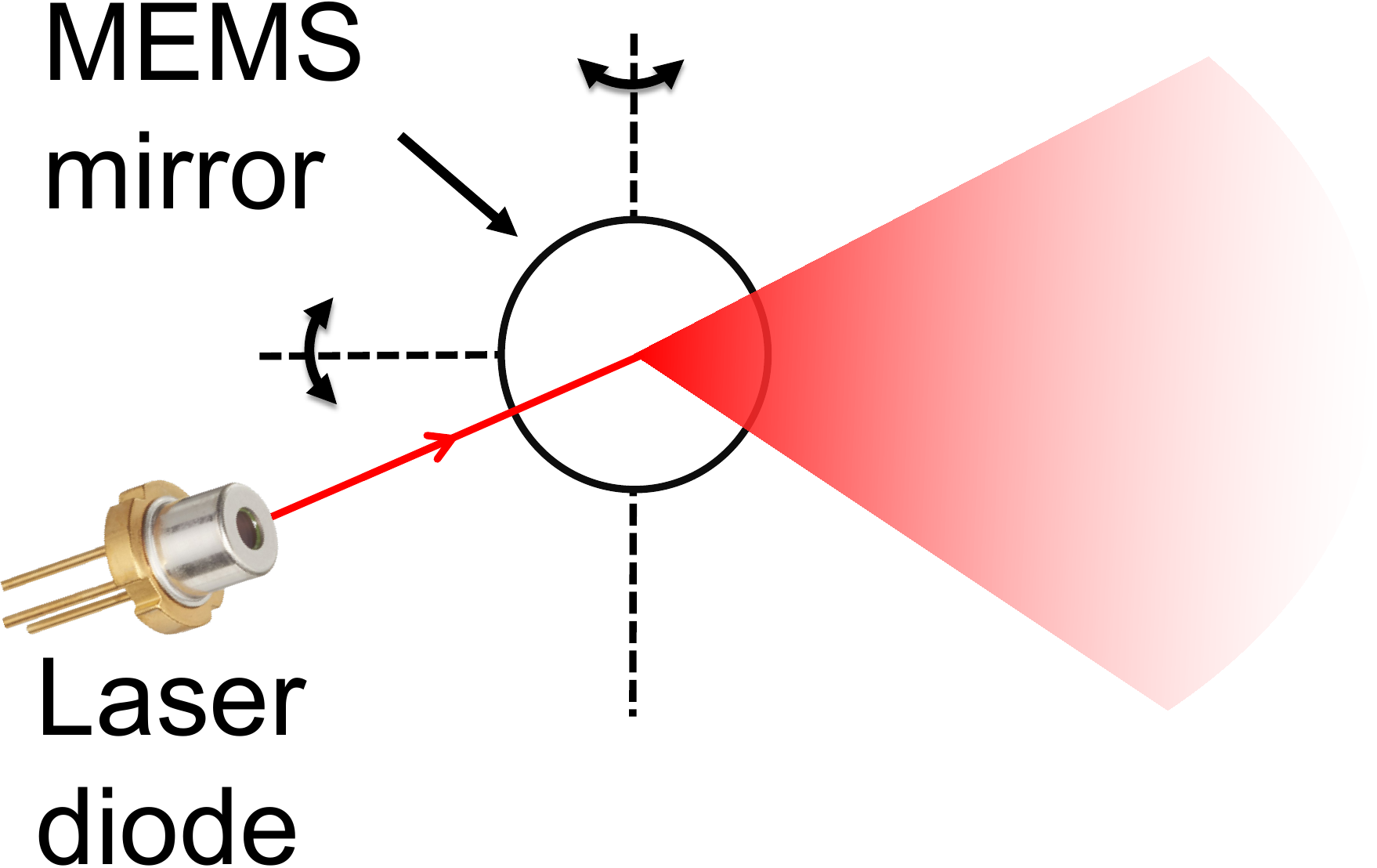}
        }
        \subcaption{Scanning principle.}
        \label{fig:lidar_scanning_principle}
    \end{subfigure}
    \caption{LiDAR measurement principles. (a) A TDC measures the ToF of emitted laser pulses for range estimation. (b) MEMS-based LiDAR steers laser beams with micro-mirrors to scan the environment.}
    \label{fig:lidar_measurement_principle}
\end{figure}

\textbf{LiDAR principle.}
As illustrated in Fig.~\ref{fig:lidar_measurement_principle}, a pulsed LiDAR measures the time-of-flight (ToF) of each laser return using a time-to-digital converter (TDC) and associates the resulting range with the beam direction, typically represented by pitch and yaw. MEMS-based sensors steer the beam continuously using micro-mirrors, producing a sequence of polar measurements that are converted to Cartesian points $(x,y,z)$. These ranging and scanning mechanisms directly give rise to the range quantization and beam-direction characteristics modeled in our synthetic LiDAR generation.

\textbf{LiDAR measurement characteristics.}
To synthesize facial point clouds representative of real vehicle-mounted LiDAR rather than idealized uniform sampling, we focus on two sensor-side characteristics that directly affect facial observations: \textit{range quantization} and \textit{beam-direction variation}. We characterize both using a RoboSense M1P, the same sensor used in our real-capture evaluation, and incorporate them into the simulator in Sec.~\ref{sec:synthetic_point_cloud}.

\begin{figure}[t]
    \begin{subfigure}[c]{0.2\linewidth}
        \centering
        \includegraphics[width=\linewidth]{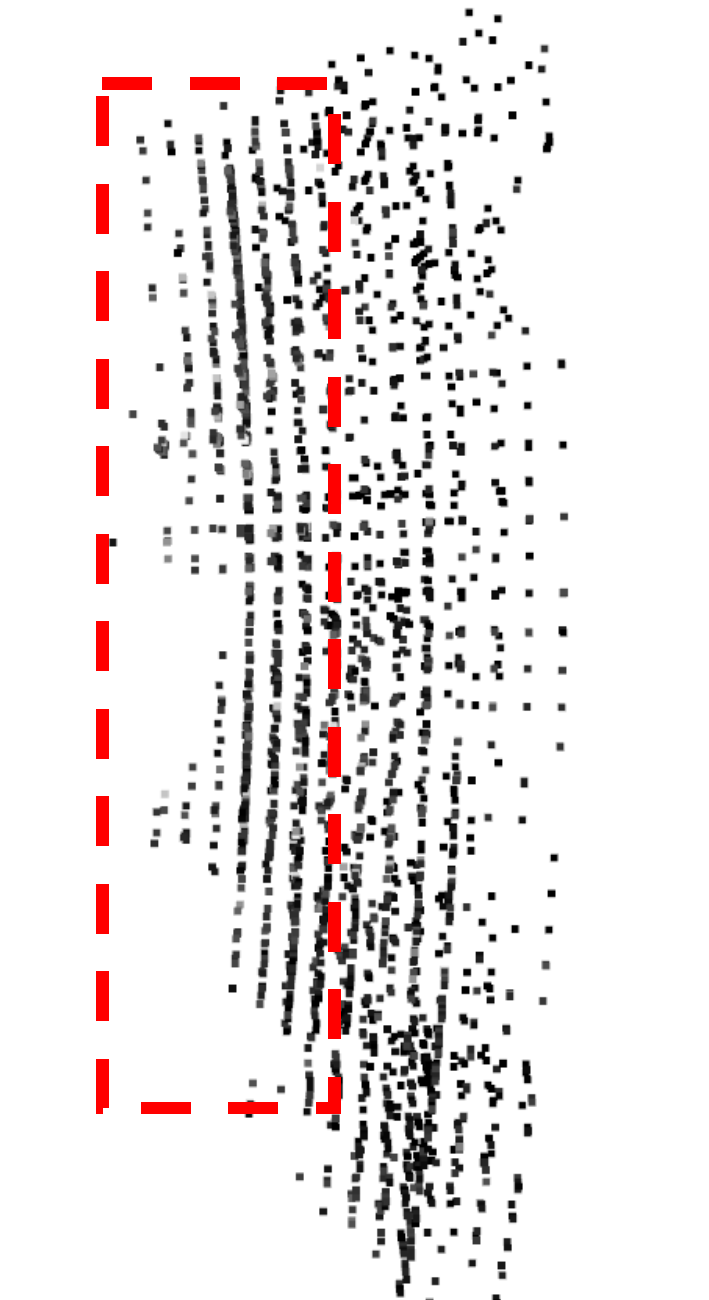}
        \subcaption{Range quantization.}
        \label{fig:lidar_range_quantization}
    \end{subfigure}
    \hfill
    \begin{subfigure}[c]{0.7\linewidth}
        \centering
        \includegraphics[width=\linewidth]{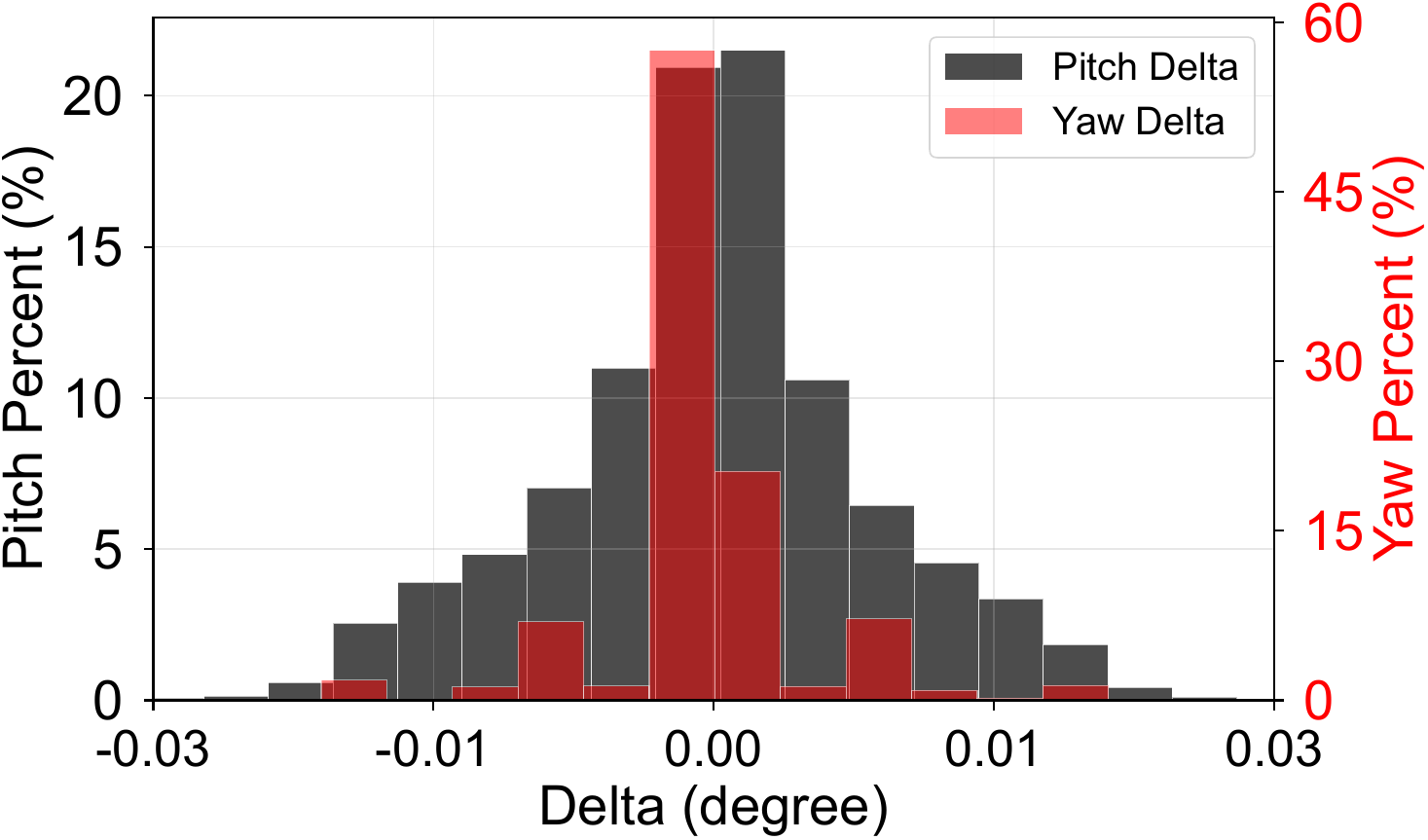}
        \subcaption{Pitch/yaw deviation distributions.}
        \label{fig:lidar_pitch_yaw_deviation}
    \end{subfigure}
    \caption{LiDAR measurement characteristics. (a) Range quantization produces discrete depth layers. (b) Measured pitch/yaw deviations of a RoboSense M1P.}
    \label{fig:lidar_distance_angle_error}
\end{figure}

\textbf{Range quantization.}
Because the TDC has finite temporal resolution, reported ranges are discretized into distance bins. The RoboSense M1P has a range resolution of 5~mm, corresponding to a quantization uncertainty of up to approximately 2.5~mm. As illustrated in Fig.~\ref{fig:lidar_distance_angle_error}(a), this discretization can produce characteristic layered structures rather than a continuous surface.

\begin{figure}[t]
    \centering
    \includegraphics[width=0.99\linewidth]{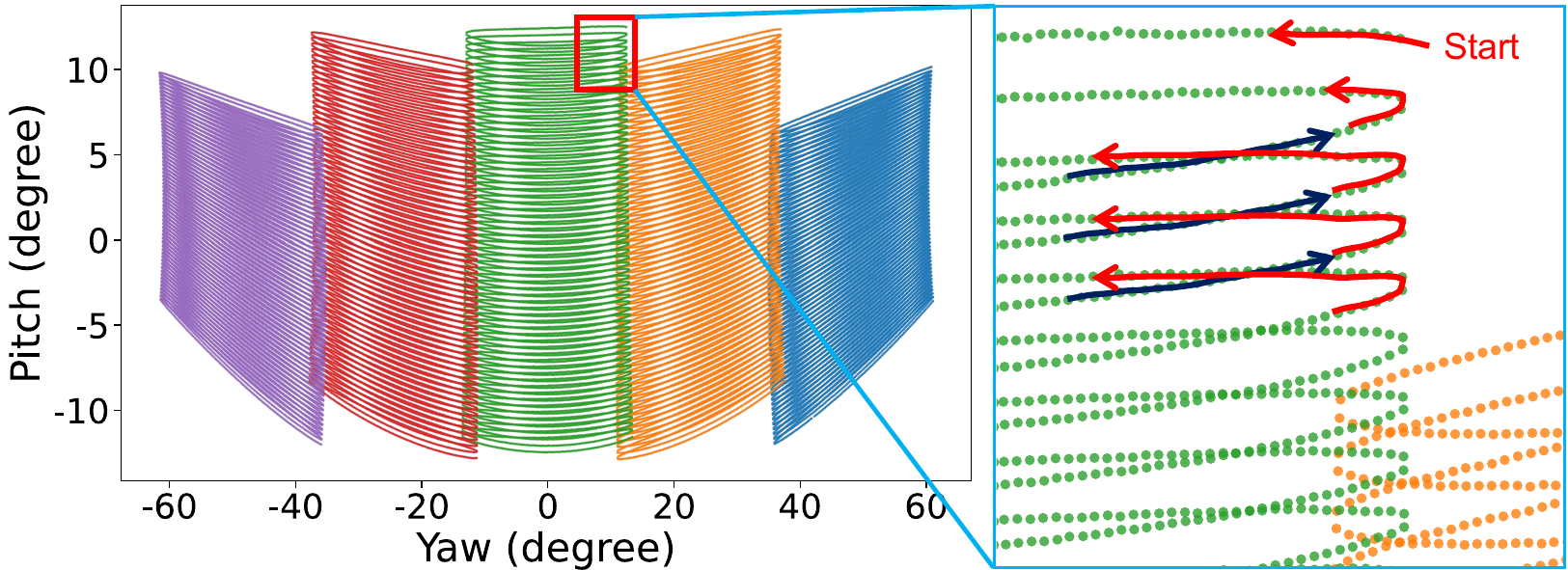}
    \caption{Empirical yaw/pitch beam pattern of the M1P. Different colors denote the five scan rings. The zoomed-in view shows the within-ring scanning order and the non-uniform, crossing scanlines produced by MEMS scanning.}
    \label{fig:empirical_beam_pattern}
\end{figure}

\textbf{Beam-direction variation.}
In a MEMS LiDAR, the effective pitch/yaw of each return depends on the mirror orientation at the firing timestamp. Timing jitter and mirror micro-dynamics therefore introduce small frame-to-frame deviations from the nominal beam direction. Measurements from our M1P show deviations within $0.02^\circ$ for both pitch and yaw (Fig.~\ref{fig:lidar_distance_angle_error}(b)). Even at the sensor's maximum detection range of 200~m, a $0.02^\circ$ deviation corresponds to a lateral offset of about 7~cm, substantially smaller than the nominal spacing implied by its $0.2^\circ$ angular resolution~\cite{M1datasheet}.

Beyond this small per-return variation, MEMS scanning also produces a characteristic non-uniform spatial sampling pattern. Fig.~\ref{fig:empirical_beam_pattern} shows the empirical yaw/pitch layout measured from real M1P point clouds, including five scan rings and locally crossing scanlines. This structure motivates the empirical beam replay used in Sec.~\ref{sec:synthetic_point_cloud}, rather than an ideal uniform angular grid.

\subsection{Face Geometry in Vehicle-mounted LiDAR}

LiDAR directly measures surface geometry through sparse 3D returns and their reflection intensity, without capturing texture or color. A substantial body of work on 3D/depth-based face recognition has shown that geometric cues alone can be identity-discriminative. However, these studies typically rely on dense and stable 3D observations acquired by dedicated 3D scanners or active depth cameras, where facial structures are well sampled~\cite{phillips2005overview}. These results establish that high-quality 3D facial geometry can support recognition, but they do not directly answer whether the same holds for vehicle-mounted LiDAR, whose facial observations are much weaker.

\begin{figure}[t]
    \centering
    \begin{subfigure}[c]{0.28\linewidth}
        \centering
        \includegraphics[width=\linewidth]{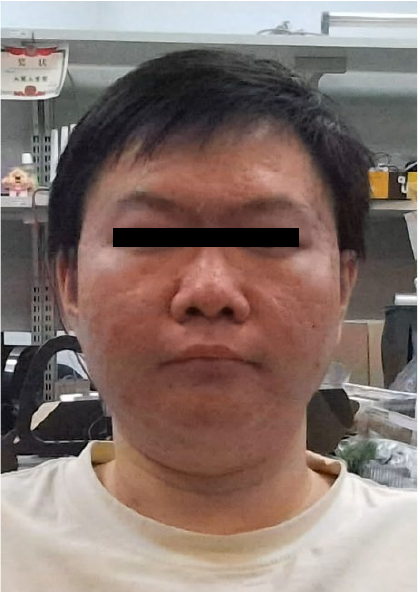}
        \subcaption{RGB image from Kinect DK.}
        \label{fig:kinect_face_point_cloud}
    \end{subfigure}
    \begin{subfigure}[c]{0.28\linewidth}
        \centering
        \includegraphics[width=\linewidth]{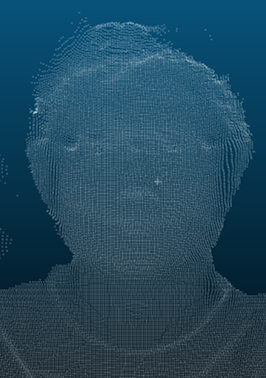}
        \subcaption{Point cloud from Kinect DK.}
        \label{fig:m1p_face_point_cloud}
    \end{subfigure}
    \begin{subfigure}[c]{0.35\linewidth}
        \centering
        \includegraphics[width=\linewidth]{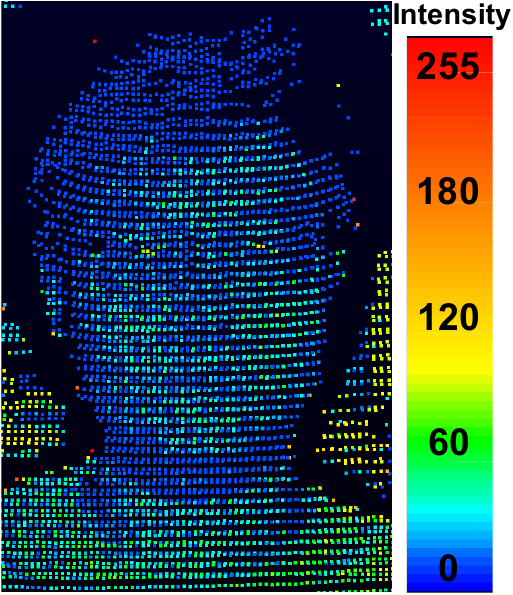}
        \subcaption{Point cloud from M1P LiDAR.}
        \label{fig:synthetic_face_point_cloud}
    \end{subfigure}
    \caption{Facial observations of the same subject from different sensing modalities: (a) RGB image, (b) dense point cloud captured by Azure Kinect DK, and (c) sparse facial point cloud captured by a real RoboSense M1P LiDAR. The M1P observation is much sparser, but still retains a recognizable face-like geometric structure.}
    \label{fig:face_geometry_examples}
\end{figure}

Fig.~\ref{fig:face_geometry_examples} provides an intuitive comparison across sensing modalities. The RGB image preserves full appearance information, and the Azure Kinect DK point cloud captures dense facial geometry. By contrast, the M1P point cloud is sparse and irregular. Nevertheless, it is not visually meaningless: the overall head contour, rough facial silhouette, and parts of the central facial region remain observable and are spatially consistent with the underlying person. Thus, the M1P scan appears as a degraded geometric observation of the same face rather than an unstructured set of points.

These properties distinguish vehicle-mounted LiDAR from the high-fidelity 3D sensing setups typically considered in prior face-recognition work. This difference arises primarily from the sensing configuration. Automotive LiDAR is designed for scene-scale perception rather than close-range face capture, so a human face subtends only a small angular extent and is sampled by relatively few beams. In addition, LiDAR acquires points through sequential beam steering instead of dense image formation, resulting in sparse and non-uniform sampling over small facial surfaces. Consequently, the resulting facial geometry is sparse, incomplete, and unstable, while still preserving non-random structure aligned with the underlying face. The key privacy question, therefore, is not whether automotive LiDAR yields high-quality 3D face data, but whether the residual facial geometry it captures remains sufficiently identity-discriminative to support identity inference or linkage.
\section{Problem Formulation and Threat Model}
\label{sec:motivation}

\subsection{Problem Formulation}
\label{sec:privacy_definition}

We study \emph{identity privacy} in vehicle-mounted LiDAR sensing: whether sparse facial geometry captured from a pedestrian can be linked to a specific individual. We define \textbf{LiDAR identity leakage} as the ability to link a pedestrian's identity from LiDAR-derived facial geometry with success meaningfully above a chance baseline under the stated threat model.

Specifically, given a LiDAR observation containing facial geometry and an external RGB face gallery, the adversary ranks candidate identities according to their similarity to the LiDAR observation. The linkage is considered successful when the true identity appears near the top of the ranked list. We focus on gallery-based linkage in which the target identity has at least one RGB reference image in the gallery; open-set recognition, where the target may be absent, is outside the scope of this work.

We consider an external RGB gallery because public face images are substantially easier to obtain than large-scale LiDAR face galleries. This setting therefore captures a practical cross-modal misuse pathway: an adversary may attempt to link LiDAR-derived facial geometry to identities using publicly available RGB references, without requiring a dedicated LiDAR biometric database. Our goal is to measure the feasibility and severity of such leakage rather than to develop a state-of-the-art face recognition system.

\subsection{Threat Model}\label{sec:threat_model}

\textbf{Adversary goal.}
The adversary aims to identify a pedestrian observed by vehicle-mounted LiDAR by linking the captured facial geometry to an identity in an external RGB gallery. The attacker produces a ranked list of candidate identities and succeeds when the correct identity is ranked sufficiently high.

\textbf{Adversary capabilities.}
We consider a low-barrier adversary relying on accessible data and commodity recognition components.

\emph{(i) RGB identity gallery.}
The adversary can collect one or more RGB face images per identity from publicly accessible sources, such as social media, online photos, or existing face datasets. The gallery may contain the target identity together with many distractor identities.

\emph{(ii) LiDAR observations.}
The adversary obtains LiDAR point clouds, or depth representations derived from them, through data access rather than active sensor manipulation. Such data may originate from shared driving datasets, debugging or incident-analysis exports, third-party data handling, or misconfigured storage. We do not assume that the adversary controls the vehicle, compromises the LiDAR hardware, or has access to identity labels associated with the captured scans.

\emph{(iii) Processing and linkage.}
The adversary can extract the facial region, project sparse LiDAR returns into a depth representation, and optionally apply depth completion before cross-modal matching. A pretrained RGB face recognizer may define the RGB embedding space, while a depth encoder can be trained or adapted using public or synthetic data to map LiDAR-derived depth into that space. Importantly, the adversary does not require paired LiDAR--RGB training data from the target identities.

We further assume that the adversary knows the LiDAR sensor type and approximate acquisition settings, which are typically available from product specifications or can be estimated from the data. We do not assume access to private biometric training data of the target individuals.

\textbf{Out of scope.}
We do not consider active manipulation of the LiDAR sensor, such as laser spoofing, or attacks that compromise the vehicle platform beyond access to collected sensor data. We also do not study inference of sensitive attributes beyond identity linkage. Our focus is passive identity leakage from facial geometry captured by vehicle-mounted LiDAR and its linkage to external RGB identity references.

\section{\name Design}\label{sec:design}

\begin{figure*}[ht]
    \centering
    \includegraphics[width=0.99\linewidth]{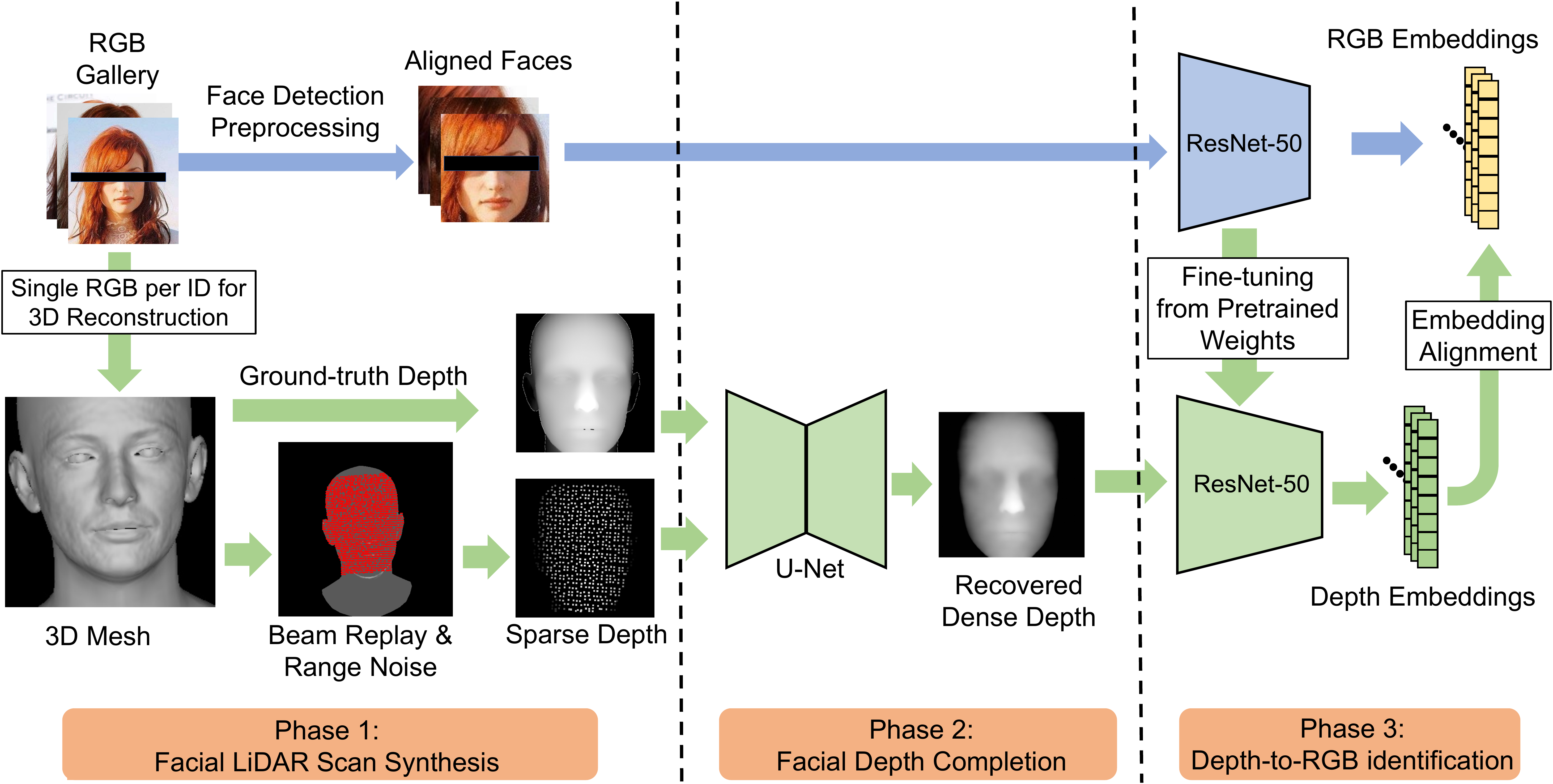}
    \caption{Overview of the \name pipeline.}
    \label{fig:system_overview}
\end{figure*}

We now present \name, a practical measurement pipeline for evaluating identity leakage from vehicle-mounted LiDAR under the threat model defined in Sec.~\ref{sec:threat_model}.

\subsection{Design Goals and Challenges}
\label{sec:design_overview}

\textbf{Design goal.}
\name is designed to measure whether sparse facial geometry captured by vehicle-mounted LiDAR can be linked to identities in an external RGB gallery. Rather than pursuing recognition performance as an end in itself, we aim for a practical and reproducible pipeline built from accessible data and standard components. Realizing this goal requires addressing three challenges.

\textbf{\textsc{C1}: Lack of face-resolved automotive LiDAR data.}
Public driving datasets rarely provide face-resolved LiDAR scans with identity labels and corresponding RGB references, while collecting such data at scale is costly and privacy-sensitive. We therefore require a scalable way to generate facial LiDAR observations without relying on idealized sampling. \name addresses this challenge through sensor-grounded synthesis based on empirical M1P beam patterns and range-measurement characteristics, followed by validation on real-person LiDAR captures.

\textbf{\textsc{C2}: Sparse and incomplete facial geometry.}
Automotive LiDAR captures substantially weaker facial geometry than dedicated 3D scanners or depth cameras: facial returns are sparse, noisy, and often incomplete (Fig.~\ref{fig:face_geometry_examples}). To examine whether densification can strengthen the identity signal available to an attacker, \name uses facial depth completion as a capability amplifier. We adopt U-Net~\cite{ronneberger2015unet} as a representative completion model; the completion network itself is not a contribution of this work.

\textbf{\textsc{C3}: LiDAR-to-RGB modality gap.}
LiDAR-derived depth encodes geometric structure, whereas RGB face images contain dense appearance information, making direct cross-modal matching difficult. At the same time, the attacker should not require paired LiDAR--RGB training data from the target identities. \name therefore uses teacher-guided embedding alignment: a frozen RGB face recognizer defines an identity-discriminative embedding space, and a depth encoder is trained to map LiDAR-derived depth representations into that space.

These challenges lead to the three core phases illustrated in Fig.~\ref{fig:system_overview}: sensor-grounded facial LiDAR synthesis (\S\ref{sec:synthetic_point_cloud}), sparse depth completion (\S\ref{sec:depth_completion}), and depth-to-RGB identification (\S\ref{sec:retrieval_attacker}). We additionally consider ID-aware re-ranking (\S\ref{sec:rerank}) as an attacker-side post-processing step to evaluate how identity-level aggregation can further amplify leakage.

\subsection{Facial LiDAR Scan Synthesis}
\label{sec:synthetic_point_cloud}

To address \textsc{C1}, Phase~1 of \name generates facial LiDAR observations at scale by separating facial geometry generation from LiDAR sensing. Starting from an RGB face image, we reconstruct a 3D facial surface and then sample this surface using sensor characteristics derived from a real RoboSense M1P. The procedure consists of four steps: 3D face reconstruction, empirical beam replay, range perturbation and quantization, and sparse depth projection.

\textbf{Step 1: Reconstructing identity-specific facial geometry.}
We obtain the facial surface for each identity through single-image 3D reconstruction. Starting from RGB face datasets including VGGFace2~\cite{cao2018vggface2}, MS-Celeb-1M~\cite{Guo2016MSCELEB1MAE}, and WebFace260M~\cite{zhu2021webface260m}, we first detect and align the face and then reconstruct a 3D mesh using DECA~\cite{DECA:yaoSiggraph2021}. DECA provides a scalable way to obtain identity-dependent facial geometry from RGB images and has been reported to achieve a median reconstruction error of 1.09~mm.

To avoid exact-image overlap between synthetic query generation and retrieval, the RGB image used for mesh reconstruction is excluded from the corresponding RGB retrieval gallery. Different images of the same identity may still appear in the gallery, as required by the identity-linkage task, but the reconstruction source image itself is never used as a retrieval reference.

\textbf{Step 2: Replaying the empirical LiDAR beam pattern.}
We next sample the reconstructed facial surface according to the spatial beam layout of the M1P. Rather than assuming a uniform angular grid or applying an artificial sampling mask, we derive a nominal beam template from thousands of real M1P frames by computing the mean yaw/pitch direction for each beam index. The resulting template preserves the characteristic non-uniform scan layout shown in Fig.~\ref{fig:empirical_beam_pattern}.

For each synthetic observation, the reconstructed mesh is placed in front of a virtual LiDAR under a sampled pose and distance. We cast one ray along each nominal beam direction and retain the intersection when the ray reaches the facial surface. The resulting point distribution is therefore determined by the measured beam layout and the geometry visible from the sampled viewpoint.

The template represents the nominal spatial sampling pattern rather than the full temporal dynamics of MEMS scanning. In particular, we do not inject frame-to-frame pitch/yaw perturbations. As shown in Fig.~\ref{fig:lidar_distance_angle_error}, the measured angular deviations are small relative to the nominal beam spacing, so we use the multi-frame mean beam directions as a stable approximation of the dominant sampling pattern.

\textbf{Step 3: Modeling range measurement effects.}
For each valid ray--surface intersection, we compute the ideal range and perturb it with
$\epsilon_r\sim\mathcal{N}(0,(2.5~\mathrm{mm})^2)$.
The perturbed range is then quantized with a 5~mm step, consistent with the reported range resolution of the M1P. The final 3D return is reconstructed from the nominal yaw/pitch direction and the perturbed, quantized range. This model captures the range-side effects considered in our simulator while keeping the angular sampling pattern fixed to the empirical beam template.

\textbf{Step 4: Projecting LiDAR returns to depth representations.}
The resulting facial returns are projected into an image-like sparse depth map $D_s$ together with a binary validity mask $M$ indicating observed pixels. From the same reconstructed mesh, we also render a dense depth map $D_{gt}$ as supervision for Phase~2. Sparse and dense depth maps are aligned to the same canonical facial template so that completion and retrieval operate in a common coordinate system.

\begin{figure}[t]
    \centering
    \includegraphics[width=0.99\linewidth]{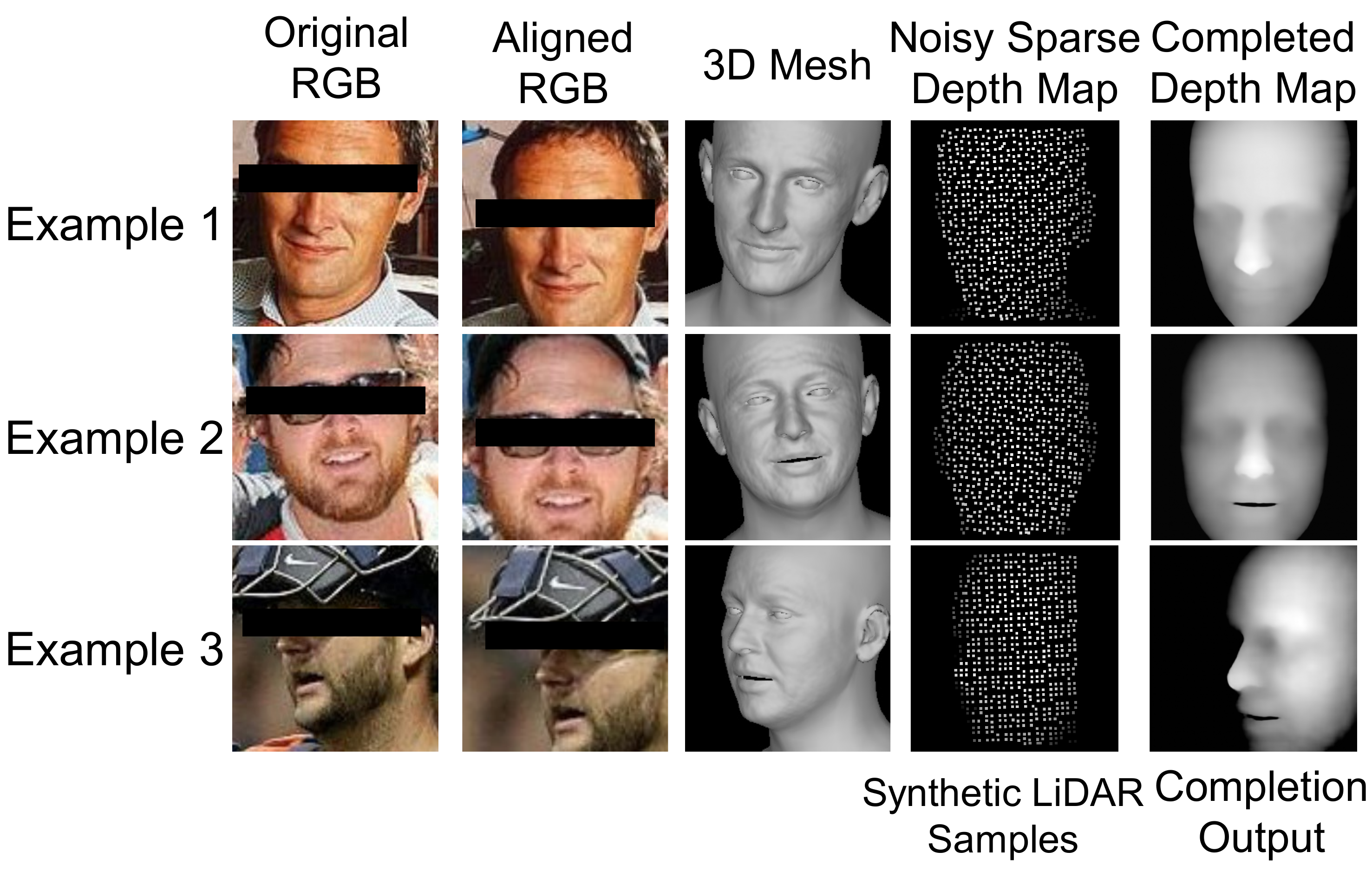}
    \caption{Representative data products across the \name pipeline, including RGB references, reconstructed facial geometry, synthetic LiDAR observations, sparse depth maps, and completed depth maps.}
    \label{fig:examples_synthesized_data}
\end{figure}

Fig.~\ref{fig:examples_synthesized_data} shows representative intermediate outputs of this process, from RGB-based reconstruction to sparse LiDAR sampling and depth completion.

\textbf{Scope of synthesis.}
The simulator is intended to reproduce selected sensing characteristics that directly determine where facial returns are sampled and how their ranges are represented; it is not a complete physical simulation of real-world LiDAR capture. Specifically, we model the empirical M1P beam layout, range perturbation, and range quantization. We do not explicitly model reflectance-dependent detection, incidence-angle effects, material-dependent dropout, hair and eyewear, motion, occlusion, background interference, or other environmental effects. The reconstructed facial surface itself also introduces RGB-to-3D reconstruction error.

Accordingly, we use the synthetic data as a controlled and scalable measurement setting rather than as a substitute for real LiDAR observations. Sec.~\ref{sec:evaluation} separately evaluates whether models trained with these synthetic observations transfer to facial scans captured by a real M1P.

\subsection{Sparse Depth Completion}
\label{sec:depth_completion}

To address \textsc{C2}, Phase~2 estimates a dense facial depth representation from the sparse depth map produced in Phase~1. We include completion to evaluate whether a denser representation of the observed facial geometry makes downstream identity linkage more effective; the completion method itself is not a contribution of this work.

\textbf{Input representation and network.}
We normalize the observed depth values using Median Absolute Deviation (MAD)~\cite{rousseeuw1993alternatives}. Let $D_s$ denote the normalized sparse depth map and $M\in\{0,1\}^{H\times W}$ its binary validity mask, where $M_i=1$ denotes an observed LiDAR return. We concatenate $D_s$ and $M$ along the channel dimension and use them as input to a U-Net-style encoder--decoder with skip connections~\cite{ronneberger2015unet}. The corresponding dense depth map $D_{gt}$ rendered in Phase~1 provides the supervision target.

\textbf{Training objective.}
Let $\hat{D}$ denote the predicted depth map and $\bar{M}=1-M$ the unobserved mask. We optimize

\begin{equation}
\mathcal{L}_{\mathrm{total}}
=
\lambda_{\mathrm{rec}}\mathcal{L}_{\mathrm{rec}}
+
\lambda_{\mathrm{grad}}\mathcal{L}_{\mathrm{grad}}
+
\lambda_{\mathrm{perc}}\mathcal{L}_{\mathrm{perc}} .
\end{equation}

Here, $\mathcal{L}_{\mathrm{rec}}$ combines masked $L_1$ and RMSE reconstruction losses over unobserved pixels, $\mathcal{L}_{\mathrm{grad}}$ penalizes differences between predicted and ground-truth depth gradients, and $\mathcal{L}_{\mathrm{perc}}$ measures higher-level structural differences using a fixed VGG feature extractor. The resulting completed depth map is used as the input to Phase~3.

We use U-Net as a representative learned completion method rather than assuming that leakage depends on a particular completion architecture. Sec.~\ref{sec:eval_rq3} further compares it with interpolation-based filling.

\subsection{Depth-to-RGB Identification}
\label{sec:retrieval_attacker}

To address \textsc{C3}, Phase~3 performs cross-modal retrieval from a completed depth query to an external RGB gallery. The key requirement is to map depth observations and RGB references into a representation in which samples of the same identity can be compared, without requiring paired LiDAR--RGB training samples from the target identities.

\textbf{Teacher-guided embedding alignment.}
We use a pretrained RGB face recognizer as a fixed teacher embedding space. The RGB branch consists of a frozen ResNet-50 face encoder that maps gallery images to identity-discriminative embeddings. The depth branch uses the same backbone architecture, initialized from the same pretrained weights, and is fine-tuned using completed depth inputs. During this training, the RGB encoder remains fixed and only the depth encoder is updated.

Let $f_{\mathrm{rgb}}(\cdot)$ and $f_d(\cdot)$ denote the RGB and depth encoders, respectively. For an RGB image $\mathbf{x}^r$ and completed depth input $\mathbf{x}^d$,

\[
\mathbf{z}^r=f_{\mathrm{rgb}}(\mathbf{x}^r),
\qquad
\mathbf{z}^d=f_d(\mathbf{x}^d).
\]

Both embeddings are $\ell_2$-normalized. At inference time, we compute cosine similarities between a depth-query embedding and all RGB gallery embeddings and rank the gallery samples accordingly.

\textbf{Retrieval-oriented training objective.}
The depth encoder is trained to increase similarity to RGB embeddings of the same identity while separating highly similar different-identity samples. We combine a multi-positive InfoNCE objective with a hard-negative margin term:

\begin{equation}
\mathcal{L}_{\mathrm{total}}
=
\mathcal{L}_{\mathrm{InfoNCE}}
+
\lambda\,\mathcal{L}_{\mathrm{hard}},
\end{equation}

where $\lambda$ controls the contribution of the hard-negative term.

\textbf{Multi-positive InfoNCE.}
Let $\mathbf{z}^d_i$ denote the $\ell_2$-normalized embedding of depth query $i$, and let
$\{\mathbf{z}^r_k\}_{k=1}^{M}$ denote the frozen RGB embeddings considered during training. We define
$s_{ik}=\mathrm{sim}(\mathbf{z}^d_i,\mathbf{z}^r_k)$
using cosine similarity and let $\mathcal{P}(i)$ contain RGB samples with the same identity as query $i$. The loss is

\begin{equation}
\small
\mathcal{L}_{\mathrm{InfoNCE}}
=
-\frac{1}{N}
\sum_{i=1}^{N}
\frac{1}{|\mathcal{P}(i)|}
\sum_{j\in\mathcal{P}(i)}
\log
\frac{
\exp\!\left(s_{ij}/\tau\right)
}{
\sum_{k=1}^{M}\exp\!\left(s_{ik}/\tau\right)
}.
\end{equation}

Unlike alignment to a single reference image, this objective treats all available same-identity RGB samples as positives, allowing the depth embedding to align with the RGB identity cluster.

\textbf{Hard-negative margin loss.}
For each depth query $i$, we select the top-$K$ most similar RGB samples belonging to different identities, denoted by $\{n^i_k\}_{k=1}^{K}$. We first compute the mean positive similarity

\begin{equation}
\small
\bar{s}^{+}_i
=
\frac{1}{|\mathcal{P}(i)|}
\sum_{j\in\mathcal{P}(i)}
\mathrm{sim}(\mathbf{z}^d_i,\mathbf{z}^r_j),
\end{equation}

and penalize hard negatives that fall within margin $m$ of this positive mean:

\begin{equation}
\small
\mathcal{L}_{\mathrm{hard}}
=
\frac{1}{N}
\sum_{i=1}^{N}
\frac{1}{K}
\sum_{k=1}^{K}
\max\!\left(
0,\,
\mathrm{sim}(\mathbf{z}^d_i,\mathbf{z}^r_{n^i_k})
-\bar{s}^{+}_i
+m
\right).
\end{equation}

The two losses therefore serve complementary roles: the multi-positive term aligns depth queries with their RGB identity cluster, while the margin term explicitly penalizes the most similar impostor samples.

\subsection{ID-aware Re-ranking for Privacy Leakage Amplification}
\label{sec:rerank}

Phase~3 produces a ranking over individual RGB images. However, the external gallery may contain multiple images of the same identity, whereas the attacker's objective is to rank candidate identities rather than individual photographs. We therefore consider an optional ID-aware re-ranking step that converts sample-level similarities into identity-level scores. This step operates only on retrieval scores and does not modify or retrain the depth-to-RGB model.

\textbf{ID-aware aggregation.}
Given a depth query $q$, we first compute its cosine similarity to every RGB gallery image and retain the top-$L$ samples. These samples are grouped by identity, after which the top-$m$ similarities associated with each candidate identity are aggregated using LogSumExp pooling:

\begin{equation}
\small
S_q(\mathrm{id})
=
\log
\sum_{g\in\mathrm{Top}\text{-}m_q(\mathrm{id})}
\exp\!\left(
\frac{s(q,g)}{\tau}
\right),
\end{equation}

where $s(q,g)$ is the cosine similarity between depth query $q$ and RGB gallery sample $g$, and $\tau$ controls the relative contribution of the retained samples.

When multiple depth observations of the same subject are available, we independently compute the identity scores for each query and average them across queries. Candidate identities are then sorted by the aggregated score to obtain an identity-level ranking. This allows the evaluation to separately measure the effect of multiple RGB references and repeated LiDAR observations on identity linkage.

\section{Implementation}\label{sec:implementation}


We implement \name in PyTorch and train the U-Net completion model on an RTX 4070 GPU and the retrieval backbones on four M40 GPUs. The implementation follows four stages: mesh reconstruction and alignment, synthetic LiDAR sampling and sparse depth projection, depth completion, and depth-to-RGB retrieval with optional re-ranking. For each identity, we reconstruct a 3D head mesh from RGB using DECA~\cite{DECA:yaoSiggraph2021} and align each depth crop to the ArcFace canonical template via a 5-point similarity transform; for VGGFace2 identities, the 5 keypoints are derived from DECA's 68 landmarks, while for real identities we manually annotate the same 5 keypoints and apply the identical alignment. We render each mesh under randomized head rotations and simulate sparse LiDAR-like returns using the per-point nominal yaw/pitch template estimated from multi-frame statistics of a real M1P sensor, then add range noise, apply range quantization, and project the resulting points into an aligned sparse depth map $D_s$ with validity mask $M$. The U-Net completion model takes $(D_s,M)$ as input and predicts a dense depth map, which is resized to the retrieval input resolution. Finally, the RGB and depth branches embed inputs into a shared space for cosine-similarity retrieval against the RGB gallery; during training, we use hard-negative mining, and at inference time we optionally apply lightweight re-ranking, including voting and ID-aware aggregation, to compute identity-level rankings. Detailed implementation settings are summarized in Table~\ref{tab:impl_settings} in the Appendix.

\section{Evaluation}
\label{sec:evaluation}

In this section, we evaluate \name to quantify identity leakage from sparse vehicle-mounted LiDAR facial scans. We prioritize real-capture evidence: we first assess whether facial LiDAR captures from a real vehicle-mounted sensor can be linked to RGB identity references, as this setting is closest to the stated threat model. We then examine whether ID-aware aggregation amplifies leakage from real captures without model retraining. We further use sensor-grounded synthetic data for large-scale and controlled analysis, where real-person collection does not readily support isolating sensing and gallery factors. Our evaluation is organized around four research questions~(RQ):

\begin{itemize}
    \item \textbf{RQ1 (Real-capture leakage):}
    \textit{Can facial scans captured by a real vehicle-mounted LiDAR be linked to RGB identity references under seen-ID and unseen-ID protocols?}
    We report real-subject retrieval results (\S\ref{sec:eval_real}).

    \item \textbf{RQ2 (Leakage amplification):}
    \textit{Can ID-aware aggregation amplify leakage from real LiDAR captures using post-processing?}
    We analyze re-ranking, Lift@$k$, and per-identity outcomes on real captures (\S\ref{sec:eval_amplification}).

    \item \textbf{RQ3 (Controlled scale):}
    \textit{How severe is identity leakage under large-scale sensor-grounded synthetic evaluation?}
    We report synthetic seen-ID and unseen-ID results, together with synthetic re-ranking amplification (\S\ref{sec:eval_synthetic}).

    \item \textbf{RQ4 (Driving factors):}
    \textit{Which sensing and gallery factors most affect leakage severity?}
    We conduct controlled studies over point density, completion choices, gallery composition, identity count, and query count (\S\ref{sec:eval_rq3}).
\end{itemize}

\subsection{Evaluation Setup and Metrics}
\label{sec:eval_setup}

\begin{table}[t]
\centering
\scriptsize
\setlength{\tabcolsep}{1.2pt}
\renewcommand{\arraystretch}{1.12}
\resizebox{\columnwidth}{!}{%
\begin{tabular}{|c|*{18}{c|}c|}
\hline
\textbf{} &
{\textbf{A}} &
{\textbf{B}} &
{\textbf{C}} &
{\textbf{D}} &
{\textbf{E}} &
{\textbf{F}} &
{\textbf{G}} &
{\textbf{H}} &
{\textbf{I}} &
{\textbf{J}} &
{\textbf{K}} &
{\textbf{L}} &
{\textbf{M}} &
{\textbf{N}} &
{\textbf{O}} &
{\textbf{P}} &
{\textbf{Q}} &
{\textbf{R}} &
\textbf{Tot.} \\
\hline

\textbf{Split} &
\multicolumn{8}{c|}{\textbf{Seen-ID}} &
\multicolumn{10}{c|}{\textbf{Unseen-ID}} &
-- \\
\hline

\textbf{Gender}         &  M &  M &  M &  M &  M &  M &  M &  M &  M &  F &  M &  F &  F &  M &  M &  M &  M &  M & -- \\ \hline
\textbf{RGB gallery}    &  7 &  9 &  9 &  7 &  9 &  9 &  9 &  9 &  9 &  9 &  9 &  9 &  9 &  9 &  9 &  9 &  9 &  9 & 158 \\ \hline
\textbf{Query (train)}  & 16 & 16 & 16 & 16 & 16 & 16 & 16 & 16 &  0 &  0 &  0 &  0 &  0 &  0 &  0 &  0 &  0 &  0 & 128 \\ \hline
\textbf{Query (unseen)} &  5 &  5 &  5 &  5 &  5 &  5 &  5 &  5 & 21 & 21 & 21 & 21 & 21 & 21 & 21 & 21 & 21 & 21 & 250 \\ \hline
\bottomrule
\end{tabular}
}
\caption{Per-identity sample counts for seen-ID training identities and unseen-ID test-only identities.}
\label{tab:counts_seen_unseen_wide_singlecol}
\end{table}

\textbf{Real-person captures.} We conduct a controlled real-subject study using facial LiDAR scans captured by a vehicle-mounted RoboSense M1P and matched RGB references. Participants were captured at close range, with LiDAR-to-subject distances varying across individuals from approximately 0.6 to 1.0~m rather than being fixed at a single distance. This variation reflects the practical difficulty of maintaining an identical face-to-sensor distance across subjects and also introduces moderate variation in facial sampling density. For each participant, we capture RGB images using a phone camera and acquire facial LiDAR scans using the same LiDAR device used to characterize the beam pattern in Sec.~\ref{sec:synthetic_point_cloud}. 

RGB embeddings are extracted with InsightFace after filtering out low-quality images, while LiDAR point clouds are processed and depth-completed using the same \name pipeline. All participants are unseen by the completion model. Per-identity sample counts are summarized in Table~\ref{tab:counts_seen_unseen_wide_singlecol}. In the seen-ID protocol, identities A--H contribute both training and query samples. In the unseen-ID protocol, the remaining identities only appear at test time. Overall, the real-person dataset contains 158 RGB reference images and 378 LiDAR depth queries with different viewing angles; the specific RGB galleries used for retrieval are defined below. 

\begin{figure}[t]
  \centering
    \includegraphics[width=0.99\linewidth]{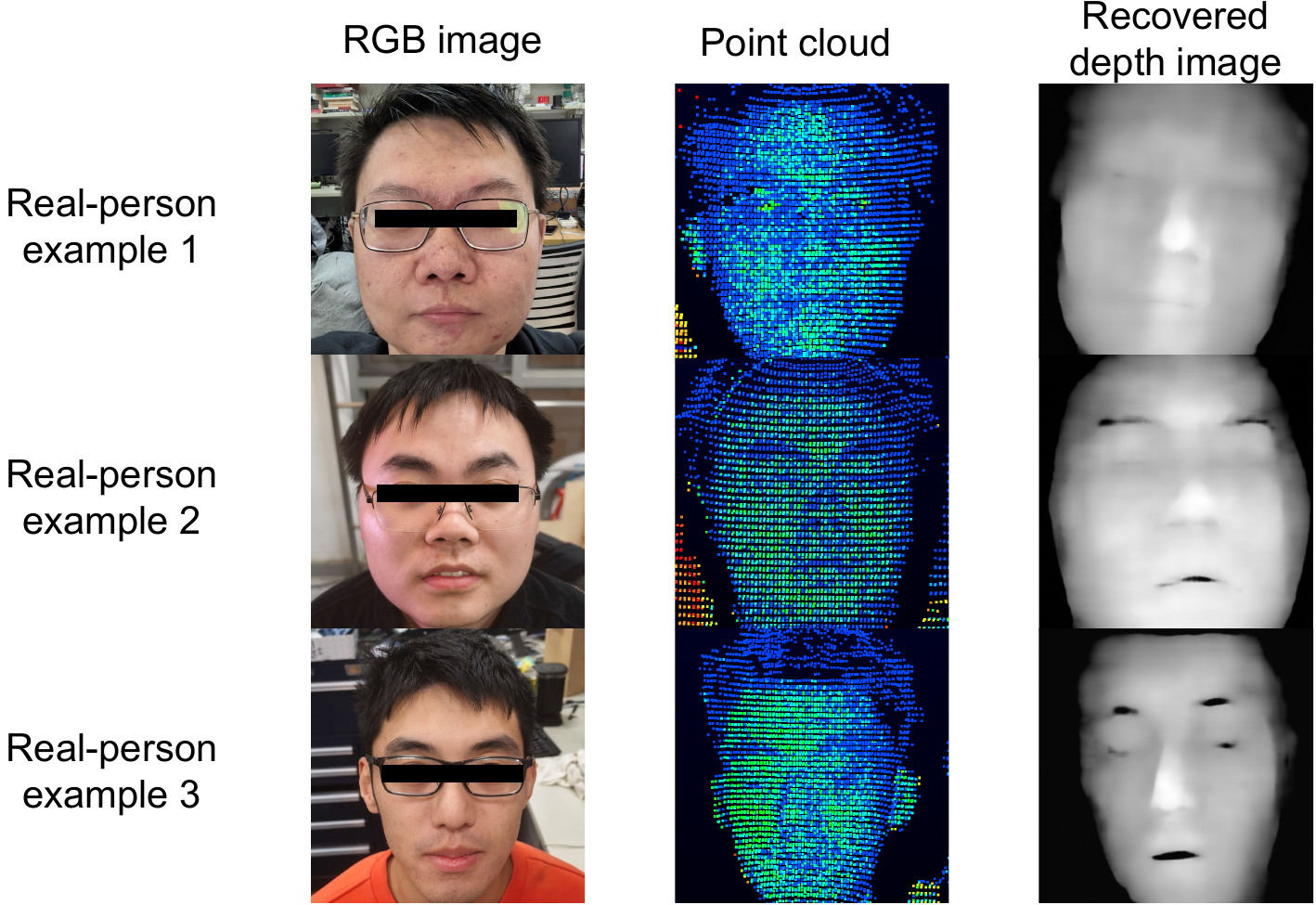}
    \caption{Representative real-person captures used in the evaluation, including RGB references, M1P point clouds, and recovered depth representations.}
    \label{fig:real_capture_examples}
\end{figure}

Fig.~\ref{fig:real_capture_examples} visualizes representative real-capture inputs and intermediate depth representations used by \name, illustrating the RGB reference, M1P point cloud, and recovered depth map for each subject. These real captures are collected under controlled conditions rather than unconstrained traffic deployment. We use them as the primary empirical check that a pipeline trained with sensor-grounded synthetic data can operate on real M1P facial scans. At the same time, we do not claim that this setup covers all deployment factors, such as motion, occlusion, background clutter, weather, or uncontrolled pedestrian pose. These factors remain important for deployment-scale risk estimation and are discussed as part of the synthetic-to-real gap.

\textbf{Synthetic data for controlled analysis.} Real-person capture provides direct empirical evidence, but it is costly to scale and does not readily support controlled factor isolation. We therefore use the sensor-grounded synthetic data from Sec.~\ref{sec:synthetic_point_cloud} for large-scale measurements and factor sweeps. Synthetic data allows us to vary point density, gallery composition, identity count, and query count while keeping other variables controlled. For all synthetic retrieval experiments, the RGB image used as the source for 3D face reconstruction is excluded from the corresponding retrieval gallery. Thus, although the gallery contains other RGB images of the same target identity, there is no exact source-image overlap between synthetic query generation and gallery retrieval.

Unless otherwise specified, we use 3000 facial points per synthetic scan as a representative operating point. This point count should not be interpreted as corresponding to a unique physical distance, because the M1P uses a non-uniform MEMS beam pattern and the number of valid facial returns also depends on pose, beam placement, and surface visibility. As a first-order geometric sanity check, however, treating the nominal $0.2^\circ$ angular resolution as an idealized two-dimensional sampling interval yields approximately 3000 samples over a face-sized region at a distance near 1~m. Thus, the default density is of the same geometric order as the farther end of our 0.6--1.0~m real-capture range, while avoiding an artificial one-to-one mapping between point count and distance. 

We do not treat synthetic results as a complete estimate of deployment-scale attack performance; rather, they complement real-capture experiments by enabling controlled and repeatable measurement at scale.

\textbf{Gallery configurations.}
Because retrieval difficulty and random-guess baselines depend on the search space, we explicitly define the RGB gallery used in each experiment. For real-person seen-ID calibration, we use two galleries: a \textbf{Real-only} gallery containing the 8 seen real identities with 68 RGB images, and a \textbf{Mixed} gallery that adds 492 synthetic distractor identities and approximately 175K RGB images. For real-person unseen-ID evaluation, the query identities are excluded from training and are searched against a mixed RGB gallery containing their real RGB references together with synthetic distractor identities; this is the gallery used for the CMC, Lift@$k$, and per-identity results in Sec.~\ref{sec:eval_real} and Sec.~\ref{sec:eval_amplification}. For synthetic large-scale evaluation, unless otherwise specified, we use a 500-ID RGB gallery with approximately 178K RGB images. In the factor analysis, we explicitly vary either the number of RGB images per identity or the number of enrolled identities to isolate the effect of gallery composition.

\textbf{Protocols.}
We use two identity protocols. In the seen-ID protocol, target identities appear during training and evaluation, representing a stronger attacker assumption and serving as an upper-bound or calibration setting. In the unseen-ID protocol, target identities are excluded from training and only appear at test time, which better matches open-world identity linkage against public RGB references. We report both sample-level ranking and identity-level ranking. \textit{Sample Rank} ranks individual RGB gallery images; success at Rank-$k$ means any correct-identity sample appears in the top-$k$. \textit{ID Rank} ranks identities after aggregating sample scores; success at Rank-$k$ means the correct identity appears in the top-$k$ identities.
\begin{itemize}
  \setlength{\itemsep}{0pt}
  \setlength{\topsep}{0pt}
  \setlength{\partopsep}{0pt}
  \setlength{\parsep}{0pt}
  \setlength{\parskip}{0pt}
  \setlength{\abovedisplayskip}{0pt}
  \setlength{\belowdisplayskip}{0pt}
  \setlength{\abovedisplayshortskip}{0pt}
  \setlength{\belowdisplayshortskip}{0pt}

  \item \textbf{Rank-$k$ and CMC.} Rank-$k$ is the fraction of queries whose correct target is within the top-$k$ list. Cumulative Match Characteristic~(CMC) plots Rank-$k$ versus $k$.
  \item \textbf{mAP@$k$.}
  For each query, AP@$k$ averages the precision values at ranks up to $k$ where a relevant gallery sample is retrieved, normalized by the smaller of $k$ and the total number of relevant samples for that query in the gallery. AP@$k$ is defined as zero when no relevant sample appears in the top-$k$. We report mAP@$k$ as the mean AP@$k$ over all queries.
  \item \textbf{Lift@k.}
  Lift@$k$ is $\mathrm{CMC}@k$ divided by the matched random-guess success rate under the same protocol, measuring how many times the attack exceeds chance. For sample-level ranking, the random baseline is defined over gallery samples. For ID-level ranking, it is defined over enrolled identities.
\end{itemize}

\subsection{Real-Person Identity Leakage}
\label{sec:eval_real}

\begin{figure}[t]
    \begin{subfigure}[t]{0.49\linewidth}
        \centering
          \includegraphics[width=\linewidth]{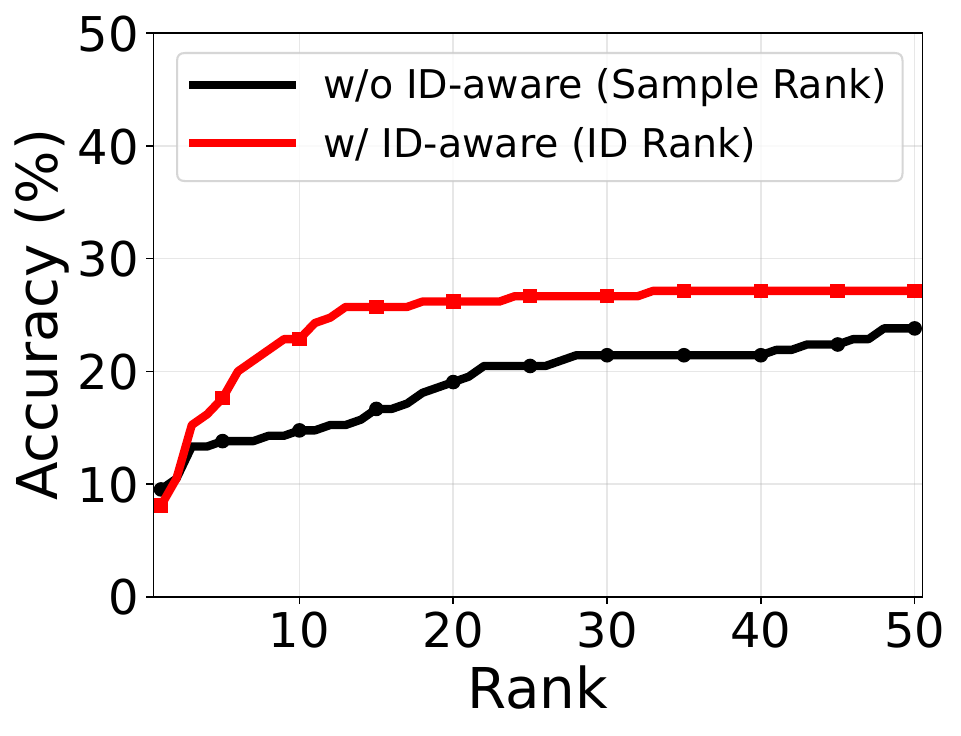}
        \subcaption{Unseen-ID CMC on real subjects.}
        \label{fig:eval_real_person_unseen_id_cmc_rerank}
    \end{subfigure}
    \hfill
    \begin{subfigure}[t]{0.49\linewidth}
        \centering
          \includegraphics[width=\linewidth]{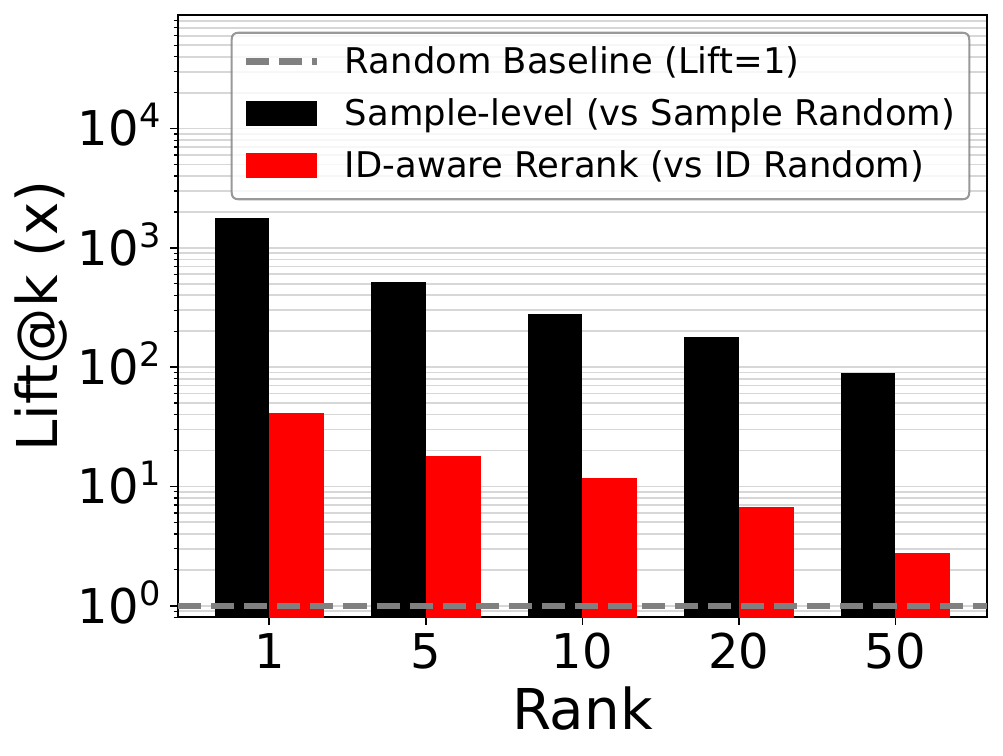}
        \subcaption{Lift@$k$ over random guessing.}
        \label{fig:eval_real_person_unseen_id_lift_rerank}
    \end{subfigure}
    \caption{Real-person unseen-ID evaluation.}
    \label{fig:eval_real_person_unseen_id_results}
\end{figure}

We first evaluate \name on real-subject LiDAR captures. This experiment assesses whether sparse facial geometry captured by a real vehicle-mounted LiDAR can be linked to RGB identity references under controlled capture conditions. We begin with the unseen-ID setting, where test identities never appear during training, and then report seen-ID results as a calibration setting for identities exposed during training.

\textbf{Unseen-ID leakage on real subjects.}
We evaluate \name on real subjects in the unseen-ID setting, where identities never appear in training and are only observed at test time. Fig.~\ref{fig:eval_real_person_unseen_id_results}(a) shows that the sample-level baseline rises gradually to about 24\% at Rank-50. This result indicates that real LiDAR-derived facial depth contains a non-trivial linkage signal even when the target identities are excluded from training. The result is modest in absolute accuracy, but it is substantially above random chance under the matched gallery setting, which is the relevant baseline for privacy leakage.

\begin{figure}[t]
    \begin{subfigure}[t]{0.49\linewidth}
        \centering
          \includegraphics[width=\linewidth]{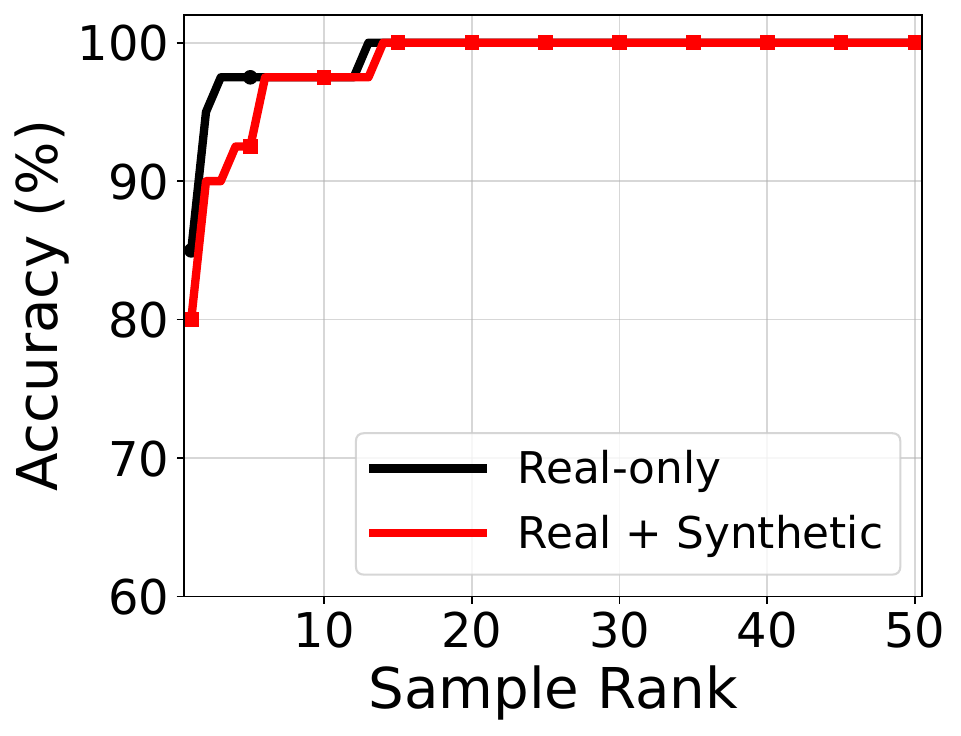}
        \subcaption{Real-person seen-ID CMC under sample-level ranking.}
        \label{fig:eval_real_person_seen_id_cmc_no_rerank}
    \end{subfigure}
    \hfill
    \begin{subfigure}[t]{0.49\linewidth}
        \centering
          \includegraphics[width=\linewidth]{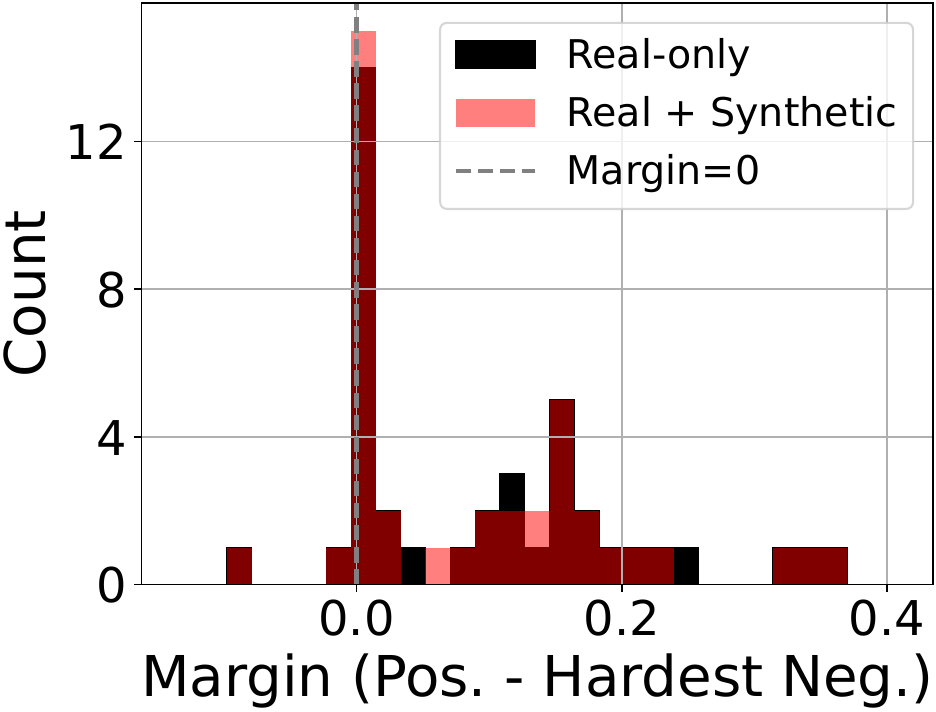}
        \subcaption{Margin distribution on real-person seen-ID.}
        \label{fig:eval_real_person_seen_id_margin_distribution}
    \end{subfigure}
    \caption{Seen-ID results on real person and margin distribution.}
    \label{fig:eval_real_person_seen_id_results}
\end{figure}

\textbf{Seen-ID calibration on real subjects.}
Fig.~\ref{fig:eval_real_person_seen_id_results} reports sample-level retrieval results under the seen-ID setting. As shown in Fig.~\ref{fig:eval_real_person_seen_id_results}(a), the CMC curves remain consistently high for both the Real-only and Mixed galleries defined in Sec.~\ref{sec:eval_setup}. Accuracy quickly saturates and approaches 100\% at moderate ranks, showing that for identities exposed during training, the learned depth-to-RGB alignment transfers to real captured depth maps and remains effective even with large-scale distractors.

To probe decision confidence, Fig.~\ref{fig:eval_real_person_seen_id_results}(b) shows the margin distribution $\Delta=s^{+}-s^{-}_{\max}$, where $s^{+}$ is the best similarity to the correct identity and $s^{-}_{\max}$ is the similarity of the hardest negative in the gallery. The Real-only and Mixed histograms largely overlap, suggesting that enlarging the gallery does not markedly shift margins for seen identities. However, both concentrate around $\Delta\approx 0$, indicating that many decisions have only a small advantage over the hardest impostor. Thus, despite near-saturated CMC, the near-zero margins reveal a limited safety buffer: additional distractors, hubness effects, or preprocessing noise may flip borderline cases.

\textbf{Findings.}
The real-person results provide direct empirical evidence that identity leakage is not limited to synthetic data. Under controlled real capture conditions, unseen identities show above-chance linkage, while the seen-ID setting confirms that the trained depth-to-RGB alignment can operate on real LiDAR depth maps. These results also motivate attacker-side aggregation, since repeated observations may stabilize weak single-query signals.

\subsection{Amplification by Re-ranking on Real Captures}
\label{sec:eval_amplification}

We next evaluate whether lightweight ID-aware re-ranking can amplify leakage from real LiDAR captures. This analysis corresponds to the optional post-processing module in Sec.~\ref{sec:rerank}. It does not retrain the model; it only aggregates the retrieval scores already produced by the depth-to-RGB identification pipeline.

\textbf{ID-aware aggregation improves real-person unseen-ID retrieval.}
Fig.~\ref{fig:eval_real_person_unseen_id_results}(a) compares sample-level retrieval with ID-aware re-ranking on real unseen identities. ID-aware re-ranking consistently improves accuracy across ranks, reaching about 23\% at Rank-10 and about 27\% at Rank-50. This indicates that identity-level aggregation across multiple queries strengthens linkage for previously unseen real subjects. In other words, an attacker can improve identification by exploiting repeated observations rather than relying on a single sparse scan.

\textbf{Leakage exceeds matched random baselines.}
Fig.~\ref{fig:eval_real_person_unseen_id_results}(b) reports Lift@$k$, computed as CMC@Rank-$k$ divided by the random-guess success probability under the matched protocol. For sample-level ranking, random guessing is defined over gallery samples; for ID-aware re-ranking, random guessing is defined over gallery identities and scales roughly with $k/|G|$. Under these protocol-matched baselines, Lift@$k$ stays well above 1 across ranks, confirming a non-trivial leakage signal for unseen real subjects. ID-aware aggregation further increases Lift@$k$, showing that lightweight attacker-side aggregation can amplify leakage without retraining.

\begin{table}[t]
\centering
\scriptsize
\setlength{\tabcolsep}{1.2pt}
\renewcommand{\arraystretch}{1.12}
\resizebox{\columnwidth}{!}{%
\begin{tabular}{|c|c|c|c|c|c|c|c|c|c|c|c|}
\hline
Setting & -- & \textbf{I} & \textbf{J} & \textbf{K} & \textbf{L} & \textbf{M} & \textbf{N} & \textbf{O} & \textbf{P} & \textbf{Q} & \textbf{R} \\
\hline

\multirow{2}{*}{\shortstack{Sample\\rank}}
& R@10 & 0.0\%  & 0.0\%  & 0.0\%  & 4.8\%  & 0.0\%  & 0.0\%  & 66.7\% & 0.0\%  & 0.0\%  & 76.2\% \\
\cline{2-12}
& R@20 & 0.0\%  & 0.0\%  & 9.5\% & 9.5\% & 0.0\%  & 4.8\%  & 71.4\% & 0.0\%  & 9.5\%  & 85.7\% \\
\hline

\multirow{2}{*}{\shortstack{ID-aware\\ID rank}}
& R@10 & 0.0\%  & 0.0\%  & 57.1\% & 19.0\% & 0.0\%  & 4.8\%  & 100\% & 0.0\%  & 0.0\%  & 100.0\% \\
\cline{2-12}
& R@20 & 0.0\%  & 0.0\%  & 85.7\% & 33.3\% & 0.0\%  & 23.8\%  & 100.0\% & 9.5\%  & 38.1\% & 100.0\% \\
\hline
\end{tabular}%
}
\caption{Per-identity real-person unseen-ID results without and with ID-aware re-ranking.}
\label{tab:real_unseen_id_detailed}
\end{table}

\textbf{Leakage is heavy-tailed across identities.}
Table~\ref{tab:real_unseen_id_detailed} provides a more security-relevant view by exposing identity-to-identity variability hidden by aggregate averages. Leakage is not uniform: several identities remain difficult with near-zero recall\footnote{For the identity that remains all 0 recall, we observed abnormal protruding artifacts in the facial point cloud that likely distort the embedding distribution. We cannot determine whether the noise is environmental or sensor-induced, so we keep these samples and report results as-is, leaving root-cause analysis to future work.}, but others become highly identifiable once multi-query evidence is aggregated. For example, identity K improves from 0.0\% to 57.1\% at R@10 and to 85.7\% at R@20; identity L increases from 4.8\% to 19.0\% at R@10 and to 33.3\% at R@20; and identity N reaches 23.8\% at R@20. Identities O and R reach 100\% recall at R@10/R@20 under ID-aware ranking. This heavy-tailed behavior matters for privacy: attackers may opportunistically succeed on easier individuals or collect additional frames and views until aggregated evidence crosses the identification threshold.

\textbf{Findings.}
Real-person leakage is amplified by simple post-processing that uses identity grouping and repeated observations. These results do not imply that all subjects are equally identifiable; instead, it shows a heavy-tailed risk profile in which some identities remain difficult while others can be reliably surfaced in a small candidate list.

\subsection{Controlled Synthetic Evaluation at Scale}
\label{sec:eval_synthetic}

Unless otherwise specified, this section uses the default 500-ID synthetic RGB gallery defined in Sec.~\ref{sec:eval_setup}. Real-person captures provide direct evidence but remain limited in scale. We therefore use sensor-grounded synthetic data to measure leakage under larger identity galleries and controlled seen-ID/unseen-ID protocols. This section does not use synthetic data as a substitute for real captures; rather, it uses synthesis as a controlled measurement tool for evaluating linkage behavior at larger scale.

\begin{figure}[t]
    \begin{subfigure}[t]{0.49\linewidth}
        \centering
          \includegraphics[width=\linewidth]{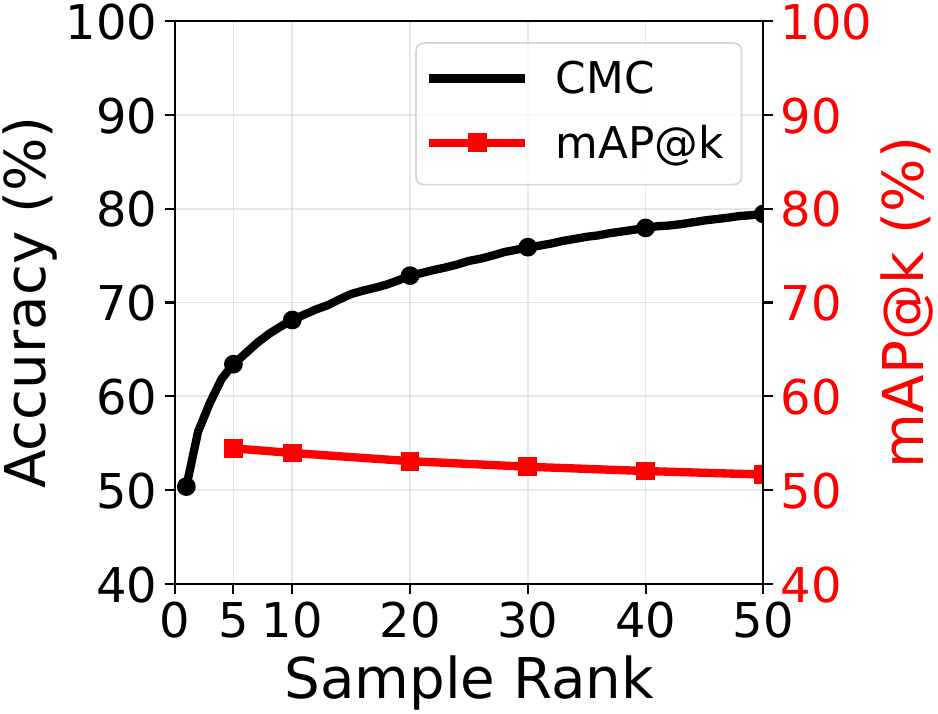}
        \subcaption{Results of seen-ID.}
        \label{fig:eval_seen_id_test}
    \end{subfigure}
    \hfill
    \begin{subfigure}[t]{0.49\linewidth}
        \centering
          \includegraphics[width=\linewidth]{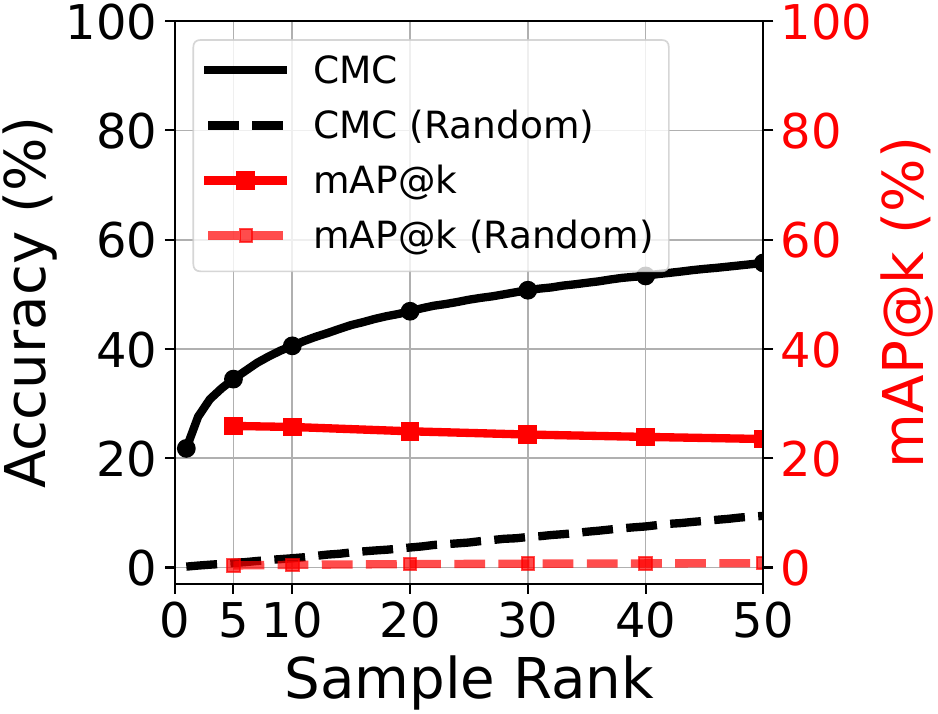}
        \subcaption{Results of unseen-ID.}
        \label{fig:eval_unseen_id_test}
    \end{subfigure}
    \caption{Controlled synthetic evaluation under seen-ID and unseen-ID protocols.}
    \label{fig:eval_results_seen_unseen_synthetic}
\end{figure}

\textbf{Synthetic seen-ID and unseen-ID.}
In the synthetic seen-ID setting (Fig.~\ref{fig:eval_results_seen_unseen_synthetic}(a)), Rank-1 reaches 50.37\% and increases to 79\% at Rank-50. This setting assumes prior exposure to target identities and should be interpreted as a stronger attacker assumption. The rising CMC curve shows that at least one correct match often appears early, while the mild decrease in mAP@$k$ indicates the ordering among multiple same-ID samples is less stable as more impostors enter the top-$k$ region.

In the synthetic unseen-ID setting (Fig.~\ref{fig:eval_results_seen_unseen_synthetic}(b)), performance increases from Rank-1 $\approx$ 21\% to near 60\% at Rank-50. The random-guess baselines under the same gallery size are far lower, and the retrieval curves remain above them across ranks. These results show that, under controlled sensor-grounded synthesis, LiDAR-derived depth retains an identity-linkage signal even for identities excluded from training.

\begin{figure}[t]
    \begin{subfigure}[t]{0.49\linewidth}
        \centering
          \includegraphics[width=\linewidth]{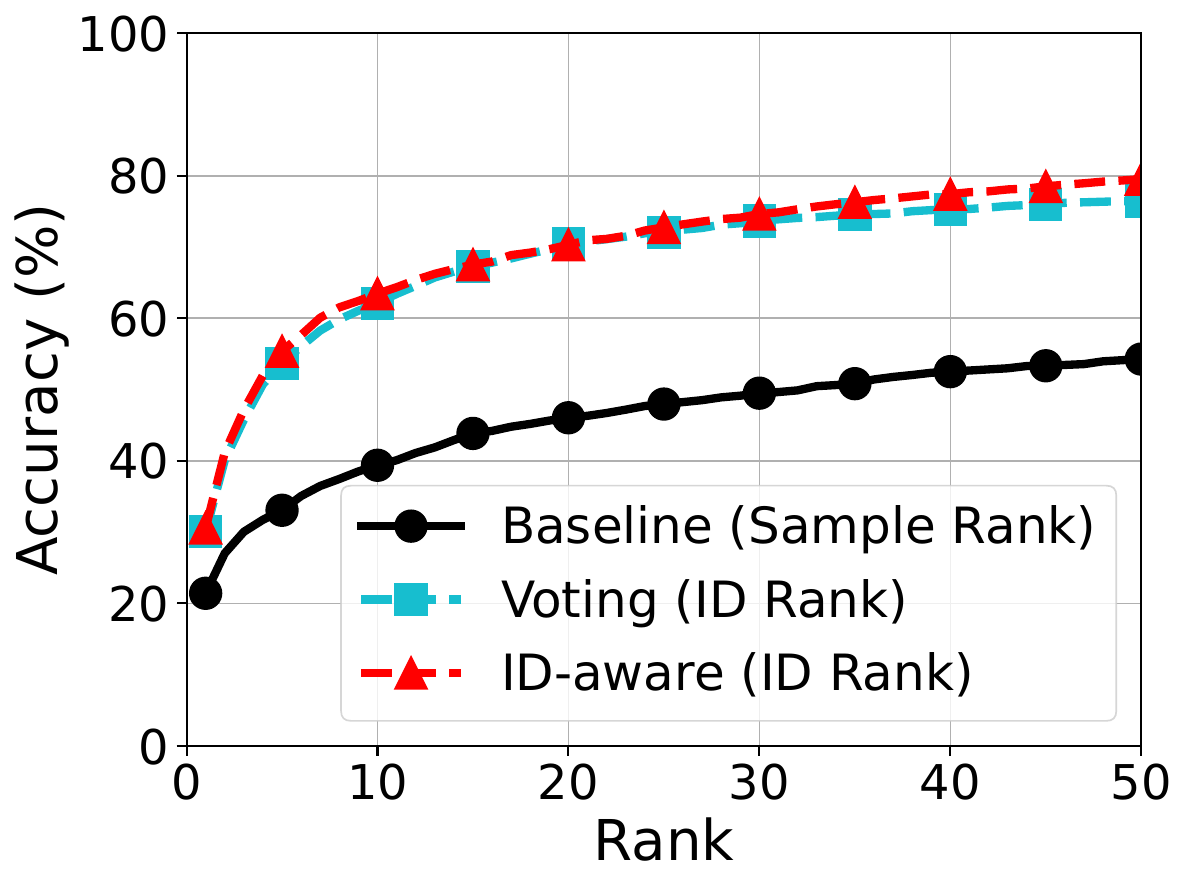}
        \subcaption{Comparison of re-ranking methods on CMC.}
        \label{fig:eval_rerank_cmc}
    \end{subfigure}
    \hfill
    \begin{subfigure}[t]{0.49\linewidth}
        \centering
          \includegraphics[width=\linewidth]{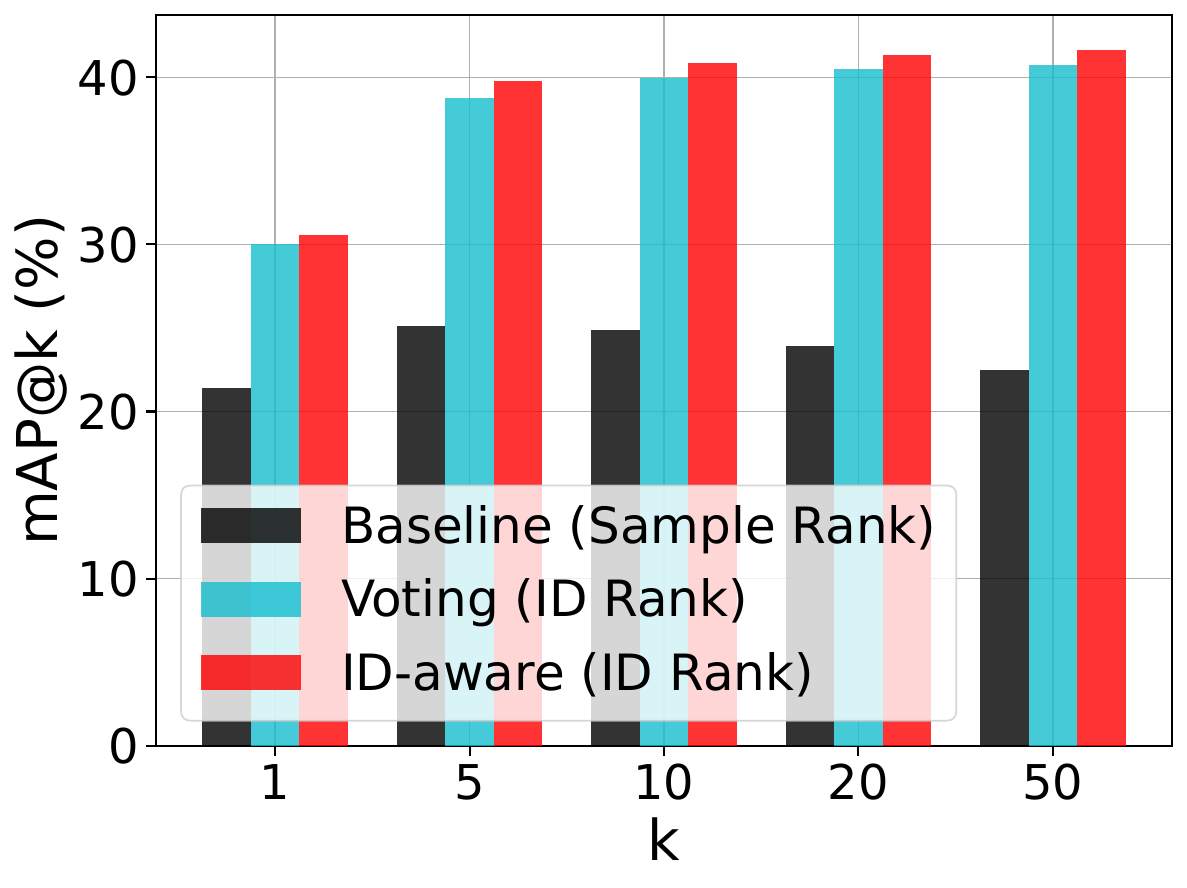}
        \subcaption{Comparison of re-ranking methods on mAP@$k$.}
        \label{fig:eval_rerank_map}
    \end{subfigure}
    \caption{Synthetic re-ranking results under controlled large-scale evaluation.}
    \label{fig:eval_results_rerank_synthetic}
\end{figure}

\begin{figure}[t]
    \centering
      \includegraphics[width=0.8\linewidth]{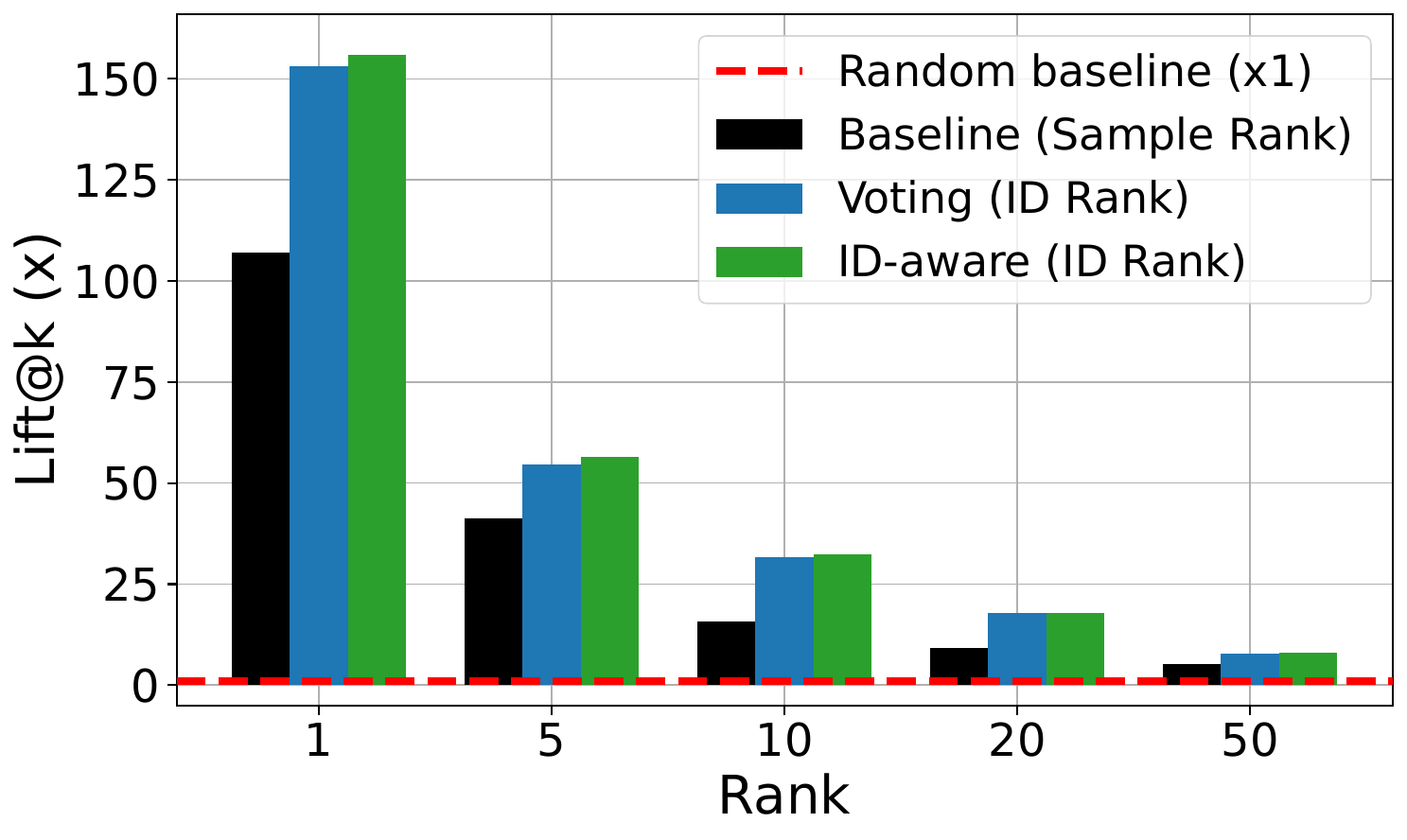}
    \caption{Synthetic Lift@$k$ over random guessing.}
    \label{fig:eval_lift_at_k}
\end{figure}

\textbf{Synthetic amplification.}
We further evaluate whether multi-query aggregation amplifies synthetic leakage at scale. Fig.~\ref{fig:eval_results_rerank_synthetic} compares the sample-level baseline with voting and ID-aware aggregation. Both strategies lift the CMC curve and improve mAP@$k$, with ID-aware aggregation performing best across ranks. This suggests that identity-level aggregation better leverages the redundancy of multi-photo galleries and suppresses isolated high-scoring impostors.

Fig.~\ref{fig:eval_lift_at_k} reports Lift@$k$ over matched random baselines. The sample-level baseline already reaches Lift@1 = 107.1$\times$, while voting increases Lift@1 to 153.2$\times$ and ID-aware aggregation to 156.0$\times$. At $k=10$ and $k=20$, amplification is nearly two-fold relative to the baseline. These controlled results complement the real-person re-ranking results by showing that the same attacker-side aggregation strategy remains effective under larger synthetic galleries.

\textbf{Findings.}
Synthetic evaluation shows that the leakage signal is not merely a small-gallery artifact: under controlled sensing assumptions, depth-to-RGB linkage remains above chance at scale and is further strengthened by ID-aware aggregation. These results should be interpreted together with the real-person experiments, which provide the primary empirical check on transfer to real LiDAR captures.

\subsection{Factors Affecting Leakage Severity}
\label{sec:eval_rq3}

Because real-person captures do not readily support controlled factor sweeps at scale, we use the sensor-grounded synthetic setting to isolate key factors. Unless otherwise specified, the following experiments use ID-aware re-ranking and report identity-level ranking~(ID Rank), where higher values indicate stronger leakage.

\begin{figure}[t]
    \begin{subfigure}[t]{0.49\linewidth}
        \centering
          \includegraphics[width=\linewidth]{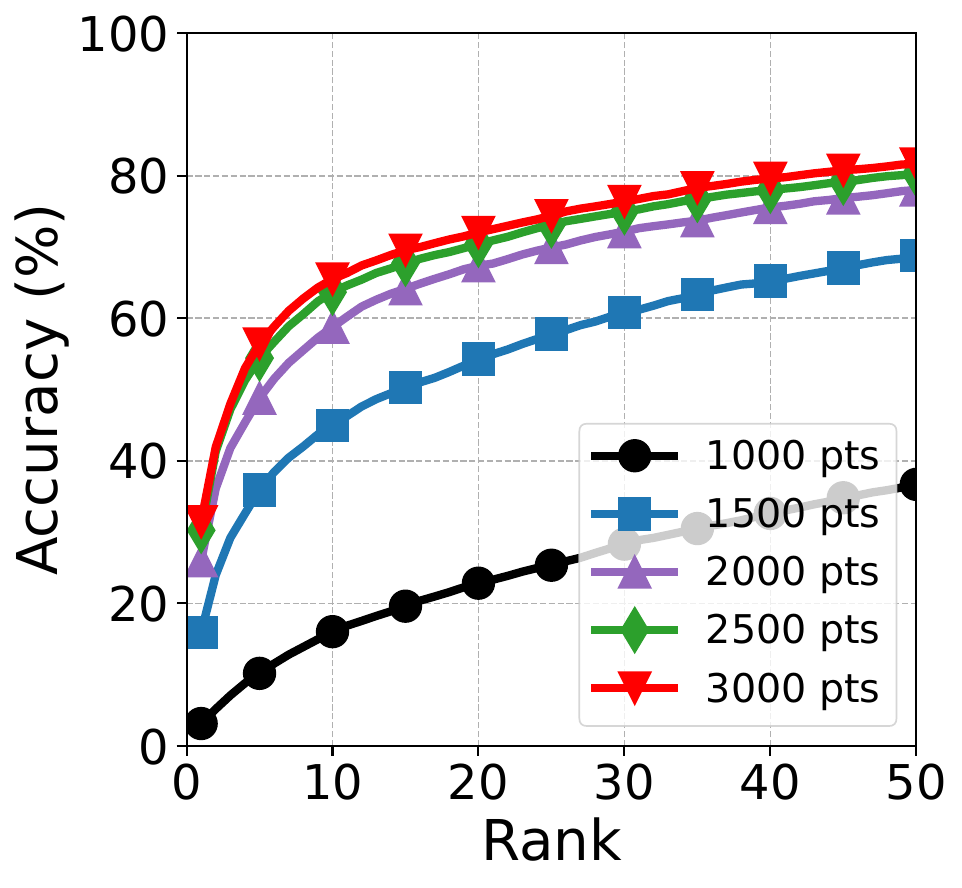}
        \subcaption{Different point density.}
        \label{fig:eval_point_density_cmc}
    \end{subfigure}
    \hfill
    \begin{subfigure}[t]{0.49\linewidth}
        \centering
          \includegraphics[width=\linewidth]{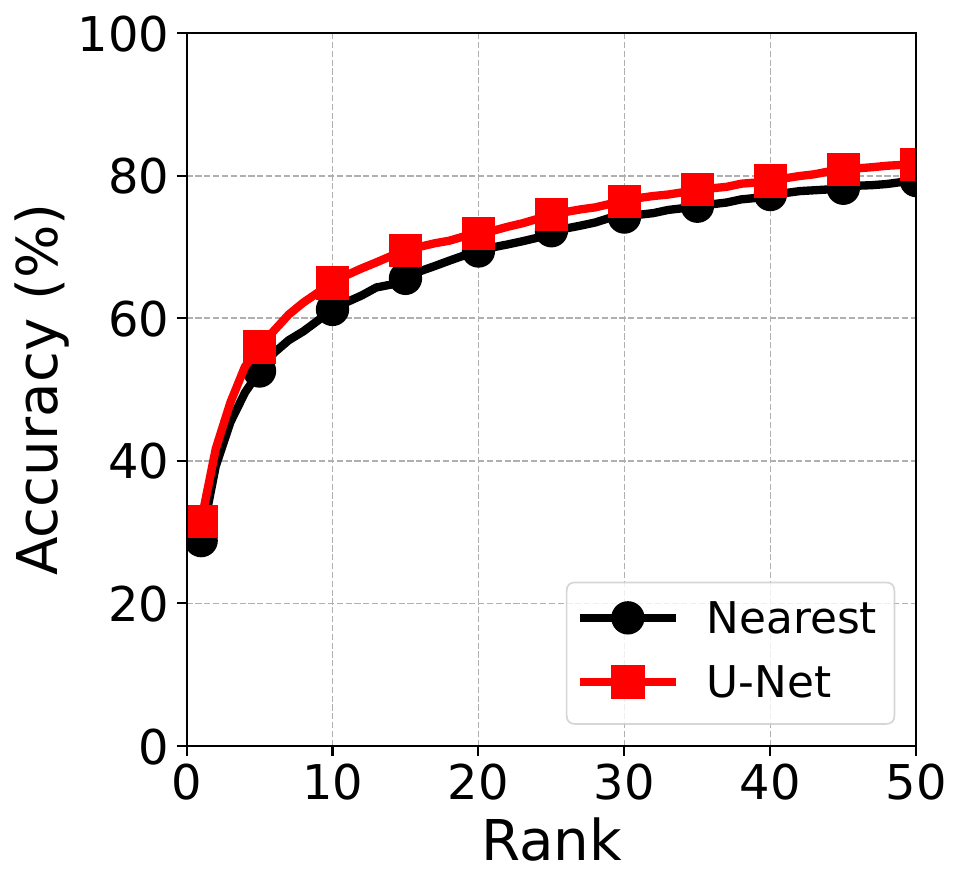}
        \subcaption{U-Net vs nearest interpolation.}
        \label{fig:eval_unet_vs_interpolation}
    \end{subfigure}
    \caption{The effect of point density and completion method.}
    \label{fig:eval_point_density_completion}
\end{figure}

\textbf{Point density.} Fig.~\ref{fig:eval_point_density_completion}(a) varies the number of retained facial points from 1000 to 3000. We treat point density as a controlled measure of the geometric evidence available to the attacker rather than as a direct proxy for physical distance. In an ideal uniform-angular model, facial point density decreases approximately with the projected angular area and therefore roughly with the inverse square of distance. The real M1P beam pattern, however, is non-uniform (Fig.~\ref{fig:empirical_beam_pattern}), and valid returns additionally depend on head pose, surface visibility, and beam placement. Consequently, different point counts can arise at the same physical distance. 

Since we resample the same underlying scans, this experiment isolates density from pose and identity effects. Leakage increases monotonically with facial point density: more points preserve more stable facial geometry, including facial contours and mid-face structures, allowing the correct identity to appear earlier in the ranking. The gain gradually saturates at higher densities, suggesting diminishing returns once salient facial structure is sufficiently sampled.

\textbf{Completion method.}
Fig.~\ref{fig:eval_point_density_completion}(b) compares U-Net completion with a simple nearest-neighbor interpolation baseline. The interpolation baseline fills missing pixels from the nearest observed depth sample within a distance threshold and treats farther pixels as background. Despite its simplicity, it follows the same trend as U-Net and achieves comparable retrieval performance. This suggests that completion amplifies leakage mainly by densifying sparse geometry, rather than by relying on model-specific hallucinated identity cues.

\begin{figure}[t]
    \begin{subfigure}[t]{0.49\linewidth}
        \centering
          \includegraphics[width=\linewidth]{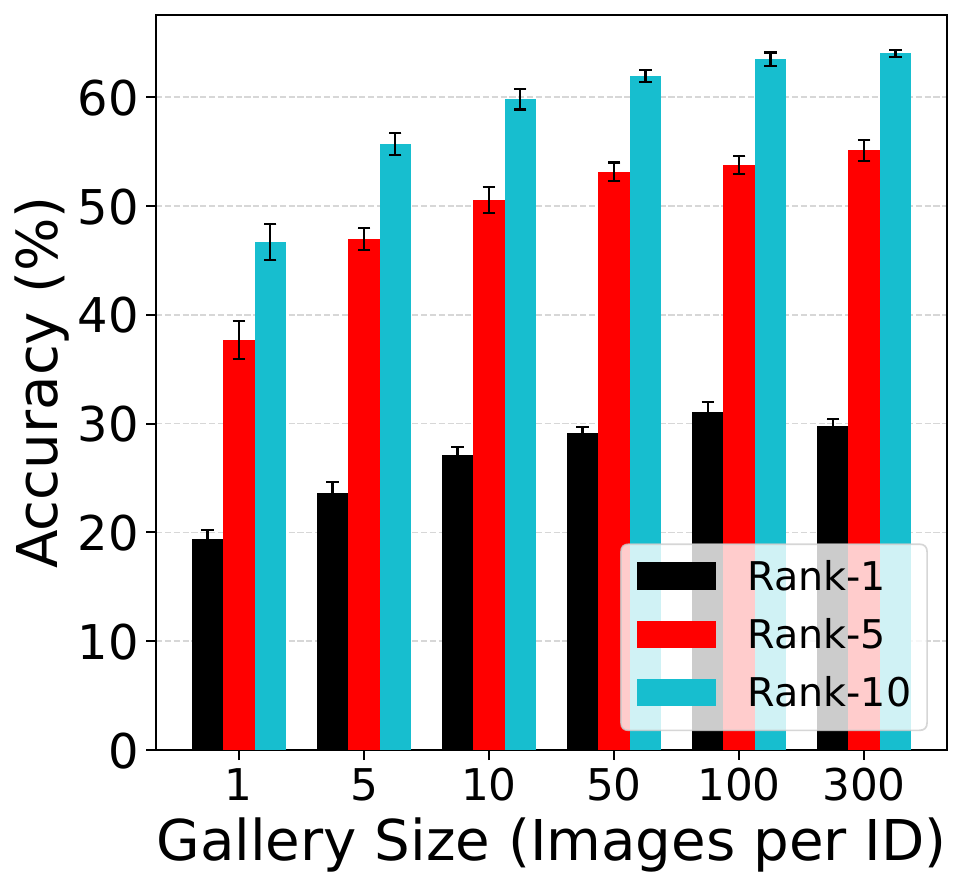}
        \subcaption{Images per ID.}
        \label{fig:eval_gallary_size}
    \end{subfigure}
    \hfill
    \begin{subfigure}[t]{0.49\linewidth}
        \centering
          \includegraphics[width=\linewidth]{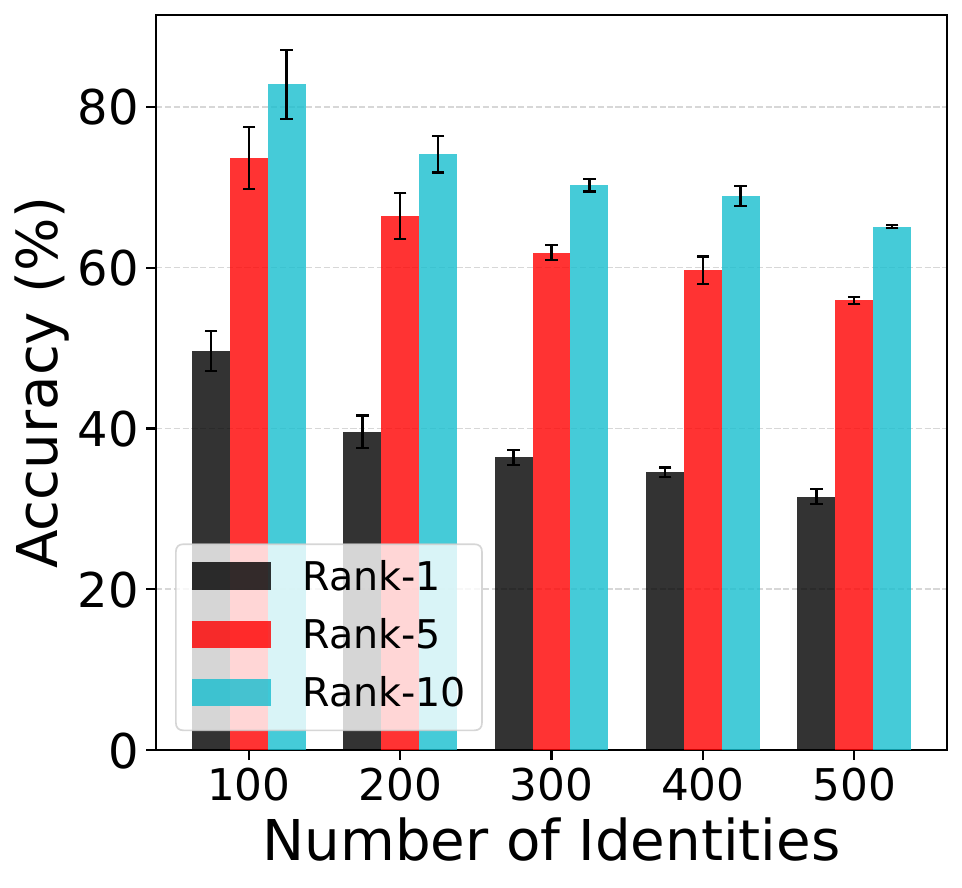}
        \subcaption{Number of identities in gallery.}
        \label{fig:eval_id_number}
    \end{subfigure}
    \caption{The effect of gallery size and identity count.}
    \label{fig:eval_gallary_id_size}
\end{figure}

\textbf{Images per identity.}
Fig.~\ref{fig:eval_gallary_id_size}(a) varies the number of RGB images per identity from 1 to 300. Rank-1/5/10 accuracy increases as the per-identity gallery becomes richer, with the largest gains from sparse to moderate coverage. More images provide pose, expression, and illumination diversity, increasing the chance of a close cross-modal match. Thus, individuals with larger public photo footprints may face higher linkage risk.

\textbf{Number of identities.}
Fig.~\ref{fig:eval_gallary_id_size}(b) increases the number of enrolled identities from 100 to 500 while keeping the per-identity image count fixed. Accuracy decreases as the gallery grows because each query faces more distractors, especially at Rank-1. However, performance remains above chance at 500 identities, indicating that a larger distractor pool reduces but does not eliminate the linkage signal.

\begin{figure}[t]
    \centering
      \includegraphics[width=0.85\linewidth]{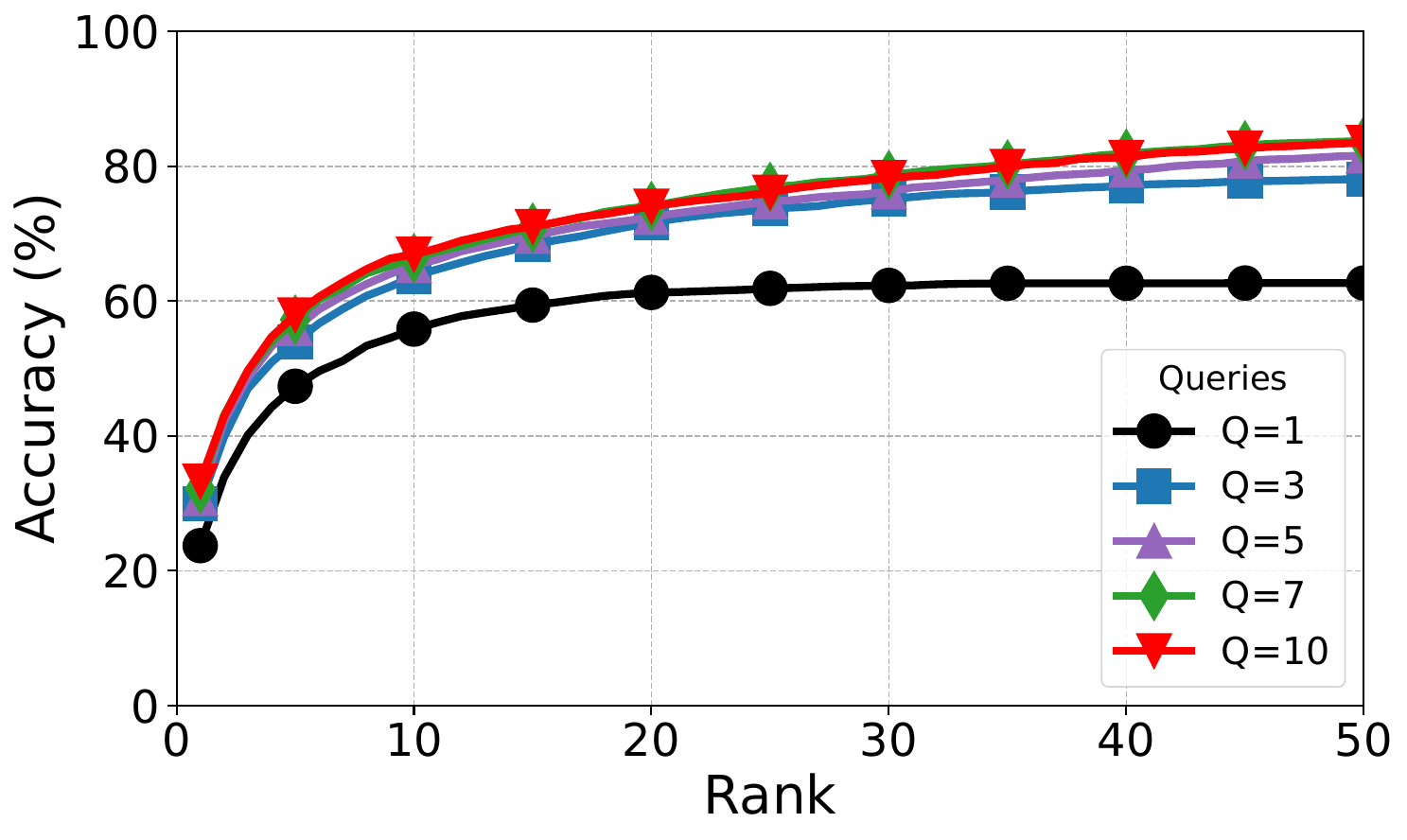}
    \caption{Query number vs.\ CMC.}
    \label{fig:eval_query_number_cmc}
\end{figure}

\textbf{Number of queries.}
Fig.~\ref{fig:eval_query_number_cmc} varies the number of depth queries $Q\in\{1,3,5,7,10\}$ from the same identity. Multi-query aggregation improves the CMC curve over the single-query setting, with the largest gain from $Q{=}1$ to $Q{=}3$ and diminishing returns beyond $Q{\ge}5$. This shows that repeated LiDAR observations can amplify leakage by stabilizing sparse and noisy evidence across scans.

\textbf{Summary.}
Overall, leakage increases with the available evidence on both sides of the linkage: denser LiDAR observations and completion strengthen the geometric signal, while richer RGB galleries and repeated queries improve cross-modal matching. Larger identity galleries make retrieval harder, but do not remove above-chance leakage under the evaluated scale.

\section{Discussion}\label{sec:discussion}

\textbf{Controlled real-world capture.} Our real-person experiments were conducted at favorable distances of approximately 0.6--1.0 m under controlled conditions. Real driving scenarios may involve greater distance, motion, pose variation, occlusion, clutter, and weather, which can substantially reduce facial returns. The reported Rank-(k) results therefore characterize identity leakage when face-resolved observations are available, rather than average identification rates for arbitrary pedestrians in traffic.

\textbf{Synthetic-to-real gap.} Our simulator models selected M1P characteristics, including beam layout, range perturbation, and quantization, but omits effects such as reflectance, motion, occlusion, material-dependent dropout, and background interference. Synthetic facial geometry is also reconstructed from RGB images and inherits reconstruction errors and model priors. The synthetic results should therefore be interpreted as controlled measurements of leakage trends rather than estimates of real-world attack performance. Larger real-world evaluations across different sensors and conditions are needed to characterize this gap.

\textbf{Threat-model boundaries.} We consider gallery-based linkage where the target identity is present in an external RGB gallery and assume access to raw or lightly processed point clouds and a usable facial region. Open-set recognition and automatic face localization and alignment from unconstrained scene-level point clouds are not evaluated. Our results therefore measure the identity information available in face-resolved LiDAR observations rather than the success rate of a fully automatic scene-to-identity attack pipeline.

\section{Related Work}
\label{sec:related_work}


\textbf{Camera-based privacy and re-identification.}
Privacy risks of vehicle-mounted cameras have been widely studied, including person re-identification, face recognition, and broader surveillance concerns in traffic scenes~\cite{hsu2020traffic}. Existing camera-based ReID methods typically extract appearance cues from in-vehicle or roadside images/videos~\cite{de2021vehicle}, relying on texture and semantic attributes such as clothing, color, facial appearance, and other fine-grained visual details~\cite{khan2023visual}. Cross-modal variants, such as RGB--IR ReID, further improve robustness when visible appearance degrades~\cite{wu2017rgb}. This line of work establishes the privacy sensitivity of imaging sensors, but remains appearance-driven and does not address whether non-imaging 3D sensors can leak identity through geometry alone.

\textbf{Radar-based person re-identification.}
Recent work has also examined identity leakage from automotive radar. Cheng \emph{et al.}~\cite{cheng2021person} study person re-identification using 77~GHz FMCW radar and show that radar returns can encode identity-discriminative cues such as micro-Doppler signatures and velocity-related features. These results suggest that non-visual sensing modalities may still expose identity information. However, radar measurements are dominated by coarse kinematics and reflectivity patterns, and do not resolve fine-grained facial geometry. Radar-based identity cues therefore differ substantially from the geometry-driven facial leakage studied in this work.

\textbf{LiDAR-based re-identification and privacy.}
The closest related work studies LiDAR-based person re-identification. Guo \emph{et al.}~\cite{guo2024lidar} collect a large-scale in-vehicle LiDAR ReID dataset and show that pedestrians can be re-identified using body-level cues such as height, body shape, and gait. Their setting focuses on full-body point clouds and recognition performance, where identity cues are dominated by global structure and motion. In contrast, our work is face-centric: we study whether sparse facial point clouds captured by vehicle-mounted LiDAR can be linked to external RGB identity references through cross-modal retrieval. This measures a complementary privacy risk that is not captured by body- or gait-based LiDAR ReID settings.

\section{Conclusion}
Vehicle-mounted LiDAR is often viewed as less privacy-invasive than cameras because it lacks photographic texture. This paper shows that this assumption is incomplete: under the stated threat model, sparse facial geometry captured by automotive LiDAR can support above-chance identity linkage against external RGB galleries. We introduced \name, a low-barrier measurement pipeline for depth-to-RGB identity linkage from LiDAR-derived facial geometry. Using real-person captures and controlled synthetic analysis, we evaluated when leakage appears and how it is amplified by factors such as point density, depth completion, RGB gallery richness, and repeated observations. Our findings suggest that raw or lightly processed LiDAR point clouds should be treated as potentially biometric data. Practical deployments should therefore consider privacy-aware retention, access control, face-region filtering, and auditing for downstream use of 3D point clouds.

\bibliographystyle{plainurl}
\bibliography{references}

\cleardoublepage
\appendix
\clearpage
\section*{Appendix}
\section{Ethical Considerations}
\label{sec:ethical_considerations}

\textbf{Data Acquisition}. 
We primarily rely on \emph{synthetic} LiDAR-like facial observations generated from 3D head models to avoid collecting large-scale real-world pedestrian LiDAR scans and to reduce risks to human subjects. For any real-person data used in the study (e.g., limited calibration or sanity-check experiments), all procedures were conducted under a protocol reviewed and approved by the Ethics Review Board at the authors' institution (Approval No. [Anonymized for Blind Review]). Such data were collected only from participants who provided informed consent after being explicitly advised that the data would be used strictly for privacy-risk/vulnerability research. We minimize the exposure of biometric information: we do not release identifiable raw real-person scans, and any examples shown in the paper are either synthetic or shared only with explicit consent for publication.

\textbf{Responsible Disclosure and Harm Minimization}. 
This work highlights a broader class of privacy leakage inherent to increasingly high-resolution 3D sensing in public spaces rather than exploiting a single vendor-specific vulnerability. Nevertheless, we take a responsible communication approach to minimize harm. We (i) clearly scope the threat model and avoid providing deployment-ready surveillance tooling, (ii) restrict shared artifacts to what is necessary for scientific evaluation (e.g., a small 10-identity pre-completion subset, two standalone notebooks for U-Net completion and depth$\rightarrow$RGB retrieval, and corresponding checkpoints), and (iii) withhold full-scale biometric data and any automation that would materially lower the barrier to misuse at scale. We also discuss actionable mitigations for system designers and practitioners, such as access control to raw point clouds, on-device filtering or obfuscation of face regions, and privacy-aware sensing configurations.

\section{Open Science}
\label{app:openscience}

The website for \name, including code, data, and pre-trained models, is available at: \url{https://anonymous.4open.science/r/lidar-face-2397}.
\subsection{Artifacts Needed to Evaluate the Core Contributions}
The following artifacts are sufficient for the program committee to verify our methodology end-to-end and reproduce the main reported metrics and figures:
\begin{itemize}
    \item \textbf{A1. Review dataset subset (10 identities, \emph{pre-completion}).}
    We provide a small review subset containing data from 10 identities \emph{before} depth completion, including the sparse/noisy depth inputs (and their validity masks) used by our completion model. This subset enables reviewers to reproduce the completion stage and subsequent depth-to-RGB retrieval on the same protocol used in the paper.
    \item \textbf{A2. U-Net completion notebook and weights.}
    \begin{itemize}
        \item \textbf{Notebook:} end-to-end inference/reproduction for sparse-to-dense completion on the 10-ID subset.
        \item \textbf{Checkpoint:} the exact U-Net weights used for the paper’s reported completion results.
    \end{itemize}
    \item \textbf{A3. Depth$\rightarrow$RGB retrieval notebook and weights.}
    \begin{itemize}
        \item \textbf{Notebook:} end-to-end evaluation for depth-to-RGB matching and identification, including CMC/Rank-$K$ and mAP@K computations.
        \item \textbf{Checkpoint:} the exact retrieval model weights used for the paper’s reported results.
    \end{itemize}
    \item \textbf{A4. Minimal evaluation scripts and protocol metadata.}
    We include configuration files, the evaluation protocol definition (gallery/query construction for the provided subset), and scripts to reproduce the core metrics.
\end{itemize}

\clearpage

\begin{table}[h!]
\centering
\small
\setlength{\tabcolsep}{4.5pt}
\renewcommand{\arraystretch}{1.15}
\caption{Implementation settings.}
\label{tab:impl_settings}
\begin{tabular}{c c}
\toprule
\textbf{Component} & \textbf{Setting} \\
\midrule
View sampling & yaw/pitch/roll $\sim \mathcal{U}[-20^\circ,20^\circ]$; $21$ views per ID \\
Points per scan & 3000 \\
Range noise & $\epsilon_r \sim \mathcal{N}(0,(2.5\text{ mm})^2)$ \\
Range quantization & 5 mm step \\
Sparse depth & $D_s$ with validity mask $M$ (template-aligned crop) \\
\midrule
Completion input & $512\times512$ \\
Completion training & 100 epochs; batch=8; Adam; lr=$1\times10^{-4}$ \\
Retrieval input & resize to $112\times112$ \\
\midrule
Embedding & 512-D, $\ell_2$-normalized; cosine similarity \\
Hard negatives & $K=100$ \\
Re-ranking & query=5; temperature=0.2 \\
\midrule
Identity splits & train IDs: $8639{=}8631{+}8$ real; test IDs: 500 \\
Depth split (seen-id) & per train-ID: 16 train / 5 held-out \\
Unseen-id queries & 10,000 depth samples total \\
Default RGB gallery & 500-ID gallery ($\sim$178K RGB images) \\
Metrics cutoff & Top-$K$ cutoff $K=100$ \\
Seeds & \{42, 2026, 3047\} (mean reported) \\
\bottomrule
\end{tabular}
\end{table}

\end{document}